\documentclass[structabstract]{aa}  
\usepackage{graphicx}
\usepackage{longtable}
\usepackage{natbib}
\bibpunct{(}{)}{;}{a}{}{,}
\usepackage{txfonts}
\newcommand{\lsim}{\
\raise-2.truept\hbox{\rlap{\hbox{$\sim$}}\raise5.truept\hbox{$<$}\ }}
\newcommand{\gsim}{\
\raise-2.truept\hbox{\rlap{\hbox{$\sim$}}\raise5.truept\hbox{$>$}\ }}
\newcommand{\reff}{$R_{eff}$~}
\newcommand{\feh}{[Fe/H]~}
\begin{document}
\title{The Star Cluster Population of the Spiral Galaxy \object{NGC
3370}\footnote{Table 3 is only available in electronic form at the CDS
via anonymous ftp to cdsarc.u.strasbg.fr (130.79.128.5) or via
http://cdsweb.u.strasbg.fr/cgi-bin/qcat?J/A+A/.}}
\author{M. Cantiello\inst{1} \and E. Brocato\inst{1} \and J. P. Blakeslee\inst{2}}
\institute{INAF-Osservatorio Astronomico di Teramo, via M. Maggini
  snc, I-64100, Teramo, Italy  \\ \email{cantiello, brocato@oa-teramo.inaf.it}
\and
Herzberg Institute of Astrophysics, National Research Council of
Canada, 5071 W. Saanich Rd, Victoria, BC  V9E 2E7, Canada \\ \email{john.blakeslee@nrc.ca}}
\date{Received -- --, --; accepted -- --, --}
 
\abstract {} {We study the photometric and structural properties of
the star cluster system in the late type Sc spiral \object{NGC 3370}.}  {BVI
observations from the Advanced Camera for Surveys on board of HST are
used to analyse in detail the colours, magnitudes and spatial
properties of cluster candidates. The final catalogue of sources used
for the study is composed by 277 objects.}  {The colour distributions
of cluster candidates appear multi-modal. Although firm age
constraints need the use of more age-sensitive indicators, the
comparison of cluster candidate colours with the colours of Galactic
and Magellanic Clouds star clusters, suggests an age difference
between the various sub-peaks, with a red old sub-system, a rich
population of intermediate age ($\sim$1 Gyr), and a blue tail of very
young (below $\sim$100 Myr) clusters. The luminosity functions appear
normal for this type of galaxy, as for the distribution of cluster
effective radii (\reff). Our analysis suggests the presence of a peak
in the \reff distribution at $\sim$3.0 pc, with blue (likely young)
cluster candidates showing smaller radii respect to red (likely old)
objects.  Finally, inspecting the properties of candidate globular
clusters, we find a colour distribution matching with Galactic
Globulars, with a median \feh$\sim-1.5$ dex, though a non negligible
tail towards lower metallicities is also present.}  {}  \maketitle
%

\section{Introduction}
The study of star clusters in galaxies provides an accurate and
relatively straightforward tool to unveil the mechanisms that produced
the present organisation and evolutionary properties of stars in the
galaxy. Star clusters allow to inspect the host galaxy both as a
system and as single star clumps hosting -relatively- simple stellar
populations. Moreover, stellar clusters are the cradle of a large
fraction of stars \citep[e.g.][and references therein]{lada03}. As a
consequence, understanding the properties of the Star Cluster (SC
hereafter) system in a galaxy helps understanding the properties of
the host galaxy. Furthermore, SCs are relatively simple stellar
systems - in first approximation single age and single metallicity
stellar populations - much simpler than the stellar system of the host
galaxies, providing the natural bridge from local single star analysis
to distant unresolved stellar population studies.

Motivated by the chance to analyse the evolutionary properties of the
host galaxy, in the last decade, especially after the HST
refurbishment, a great deal of work has been dedicated to the study
the properties (luminosity, radius, mass, etc.) of the SC system in
spiral and irregular galaxies \citep[e.g.][just to quote few
examples]{whitmore95,whitmore99,larsen00,bik03,degrijs03,gieles06a,gieles06b}.

In this work we present a study of the SC system in the spiral galaxy
\object{NGC 3370}, based on archival multi-band images taken with the HST
Advanced Camera for Surveys.

The galaxy is a nearly face on Sc spiral \citep[Hubble type
SA(s)c,][]{devaucouleurs91}, showing an intricate tangle of arms, and
mass similar to our own Milky Way. \object{NGC 3370} was the host of the type
Ia Supernova SN 1994ae \citep{sn94ae}. The light-curve of this SN Ia
is one of the nine templates used by \citet{riess96} to introduce the
SNe Ia multicolour light-curve shapes method to estimate distances.
With the advent of the high efficiency ACS camera on board of HST, and
thanks to the specific properties of the galaxy (face on, relatively
nearby spiral), \object{NGC 3370} has been targeted by an HST program to study
the light-curves of Cepheids, to provide a direct calibration of the
SN Ia distance \citep{riess05}. The program, carried out on optical V
and I images, has been integrated by complementary B band exposures
for the Hubble Heritage program.

The optical BVI data available offer an excellent opportunity to
investigate accurately not only the photometric, but also the
structural properties of the SC systems in this spiral, thanks to the
high resolution and small PSF FWHM of the ACS.

The paper is organised as follows. In section 2 we present the
observational dataset used, the data analysis carried out, and the
detection and selection criteria chosen for SC candidates. The study
of photometric and spatial properties and the comparison with models
is discussed in Section 3. We present our conclusions in Section 4.

\begin{figure*}
\centering
\caption{ACS B, V and I composite image of \object{NGC 3370}, the image size is
roughly  $3.5\times 2.5 $ arcminutes (Credits: NASA,
The Hubble Heritage Team and A. Riess - STScI).}
\label{ngc3370}
\end{figure*}

\begin{table}
\caption{NGC\,3370 properties and observations}
\begin{tabular}{l l}
\hline \hline
RA  \& Dec  (J2000)            & 0h47m04.0s +17d16m25s        \\
Morphological Classification   & Sc$^{\mathrm{a}}$            \\
Morphological T-type           & $5.0\pm1.1^{\mathrm{a}}$     \\
Total corrected B magnitude    & $\sim 11.7^{\mathrm{a}}$     \\
E(B-V)                         & 0.031$^{\mathrm{b}}$         \\
Distance Modulus               & 32.29 $\pm$ 0.06$^{\mathrm{c}}$             \\
F435W Exposure time (s)        & 9600                         \\
F555W Exposure time (s)        & 57600                        \\
F814W Exposure time (s)        & 24000                        \\
\hline
\end{tabular}
\begin{list}{}{}
\item[$^{\mathrm{a}}$] Data from Hyperleda (http://leda.univ-lyon1.fr/)
\item[$^{\mathrm{b}}$] Extinction from \citet{sfd98}
\item[$^{\mathrm{c}}$] \citet{riess05}
\end{list}
\label{tab1}
\end{table}

\section{Observations, Data Analysis and Star Cluster selections}

The ACS/WFC camera observations of \object{NGC 3370} used in this work come
from two different proposals. The proposal ID \#9351 (P.I. A. Riess)
consists of very deep F555W ($\sim$ V) and F814W band ($\sim$ I)
observations to study the Period-Luminosity relation of the Cepheids
in the galaxy \citep{riess05}, with the purpose of deriving the
distance of \object{NGC 3370} and provide a precise luminosity calibration of
the Type Ia supernova observed in the galaxy. The study realised by
\citeauthor{riess05} provided the farthest direct measurement of
Cepheids. The other set of F435W band ($\sim$ B) images was
obtained for the Hubble Heritage (proposal ID \#9696, P.I. K. Noll). A
colour image of the galaxy is shown in Figure \ref{ngc3370}. The main
properties of \object{NGC 3370}, and of the observational dataset used are
reviewed in Table \ref{tab1}.  For the distance modulus, we adopt
\citeauthor{riess05} value, $\mu_0=32.29\pm0.06$, derived from an
empirical calibration of Cepheids' Period-Luminosity relation, and an
LMC distance modulus of 18.50 mag.

Standard reduction was carried out on raw images using the APSIS image
processing software \citep{blake03}. Since ACS/WFC images are
under-sampled for wavelengths at $\lambda \lsim 11,000$ \AA, and
thanks to the fine dithering of the input images, a resolution better
than the ACS detector pixels size, 0.05\arcsec/pixel, can be achieved
with these images. We adopted a resampling of 0.035\arcsec/pixel as
input parameter for APSIS, which corresponds to $\sim 5$ pc/pixel
at the galaxy distance.

\begin{figure}
\includegraphics[width=\linewidth]{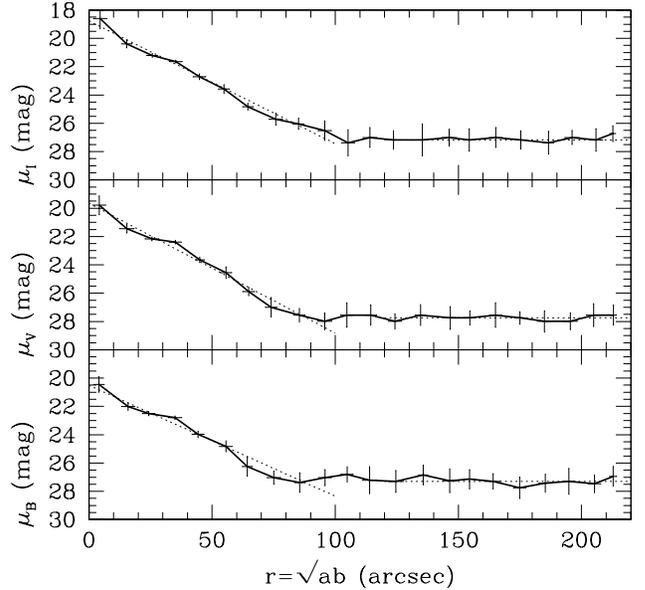}
\caption{Upper to lower panel: radial I, V and B band surface
brightness profiles of \object{NGC 3370}. The linear fit to the data at
r$\lsim$ 100\arcsec is shown with a dotted line. The horizontal dotted
line at r$\geq$120\arcsec shows the background level in the three
observed bands.}
\label{sbprof}
\end{figure}

After cosmic--ray rejection, alignment, drizzling, and final image
combination, we derived the sky background from the median counts in
the image corner with the lowest number of counts. This assumption is
supported by the fact that the surface brightness profile of the
galaxy is practically constant for
r$\gsim$120\arcsec~(Fig. \ref{sbprof}).

The source detection was obtained using SExtractor \citep{bertin96}
independently on the V and I frames. Due to the low S/N of B-band
images, the source detection in this band was done using SExtractor in
the pixel association mode (\verb ASSOC ~mode), adopting as reference
the coordinates of sources detected in the I-band frame. We used a
minimum of 5 connected pixels and 2$\sigma$ threshold for sources
detection. This resulted in a list of $\sim$20,000 objects. The
coordinates of sources detected from V and I frames were
cross-checked, allowing a maximum wandering of 4 pixels\footnote{The
alignment between single HST frames is generally better than the
maximum wandering adopted here. Our choice has been motivated by the
large spatial variability of \object{NGC 3370} surface brightness profile, and
its wavelength dependence, which may lead to different estimates of
the source centroid in different bands. A posteriori, we verified that
for none of the SC candidates in the final sample the coordinate
wandering between V and I frames exceeds $\sim$2 pixels.}, then we
performed a first very bland selection of SC candidates based on the
SExtractor parameters star/galaxy classification, object
major-to-minor axis ratio, and FWHM (object with CLASS\_STAR $=0.00$,
elongation A\_IMAGE/B\_IMAGE $\geq$ 2.5 and FWHM$>$12 were rejected).
Moreover, we rejected the sources fainter than $m_I$=27.5, and
m$_V$=28.5 mag\footnote{The aim of the selection based on the
SExtractor morphological parameters CLASS\_STAR, and A\_IMAGE/B\_IMAGE
is to remove from the candidate list all obvious non SCs. Visual
inspection of objects rejected in this way revealed that they are
mostly extended, irregular, diffuse or elongated objects. Moreover, in
a further selection based on ishape output parameters (see end of this
section), we rejected candidates with elongation estimated by ishape
$\geq$2. The limiting magnitudes adopted here are, instead, related to
the S/N ratio of ishape. Tests made on sources fainter than the
limiting magnitudes adopted showed that the S/N is in all cases much
lower than the minimum acceptable value $\sim$30 suggested by
\citet{larsen99}.}.  Using such limits we ended up with a list of
$\sim$8000 SC candidates whose structural properties were then
analysed.

\begin{figure}
\includegraphics[width=\linewidth]{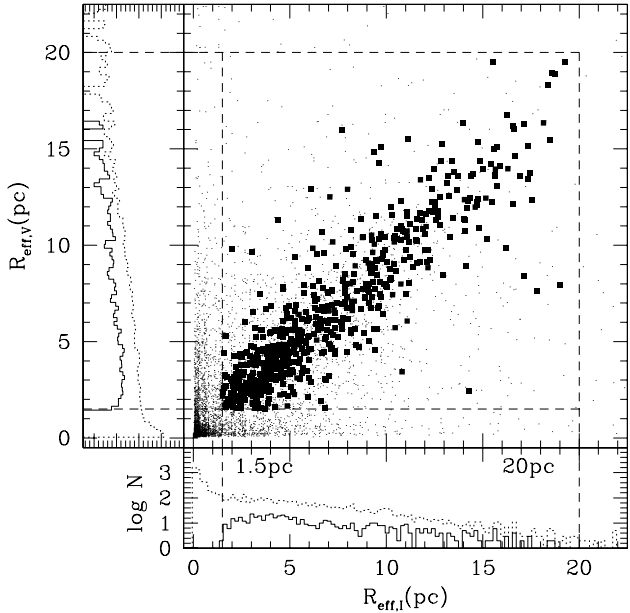}
\caption{Central panel: effective radii from the I-band frame versus
the V-band \reff (dots). The full squares show cluster candidates
whose \reff agree within $2\sigma$ in both bands (see text for
details). The lower and left panels show the histograms of radii for
the whole sample of sources (dotted lines), and for the SC candidates
with matching radii (solid line). The dashed lines show the 1.5 pc and
20 pc cut radii.}
\label{vi_ishape}
\end{figure}

Given the linear pixel scale of these \object{NGC 3370} images, and the ACS
Point Spread Function FWHM of $\sim$2.5 pixels, we are in the
condition of using the \verb ishape ~software \citep{larsen99} to
estimate the spatial extension of SC candidates. This software package
allows to determine the physical characteristics of SC candidates
(\reff, elongation, etc.) by convolving analytic profiles with the
PSF, then best-fit parameters are obtained by the $\chi^2$
minimisation. Ishape provides accurate intrinsic FWHM values of
slightly resolved objects down to 1/10 the FWHM of the PSF,
corresponding to $\sim$1.2 pc in our case. In other words, provided
that a good PSF sampling can be obtained, effective radii of star
clusters $\gsim 1.2$ pc can be retrieved. The input PSF to be
used for \verb ishape ~needs to be subsampled by a factor 10 with
respect to the input science images. As suggested by the \verb ishape ~handbook, 
we generated our PSF using the SEEPSF task of DAOPHOT within
IRAF\footnote{IRAF is distributed by the National Optical Astronomy
Observatories, which are operated by the Association of Universities
for Research in Astronomy, Inc., under cooperative agreement with the
National Science Foundation.}. To avoid contamination from clusters,
the list of stars to determine the PSF was selected on the basis of
the FWHM, colours and magnitudes from DAOPHOT. We finally used 10
bright stars (all common to the catalogue of comparison stars in Table
3 of Riess et al.), evenly distributed on \object{NGC 3370}, and all at
distances $>50 \arcsec$ from the centre of the galaxy. It is useful to
anticipate here that {\it a posteriori} all the objects used as
template for the PSF, and {\it all} the Cepheids detected by
\citeauthor{riess05} were catalogued as stars by \verb ishape .

Among the various profiles available with \verb ishape, we choose a
Moffat profile with a power law index -2.5, since it gave generally
better residuals compared to other choices.  A Moffat profile has
already been shown to be effective in fitting the profiles of both
young and old star clusters, though with slightly smaller exponents
with respect to our choice \citep[see,
e.g.,][]{elson87,larsen99,mackey03}. As discussed in \citet{larsen99},
the details of the model profile chosen are not an issue as long as
the the \reff values are derived for sources with intrinsic sizes
smaller or similar to that of the PSF, as in the present case. The
analysis was carried out independently on the I, V, and B frames,
although only I and V band data were finally used to estimate the
average effective radii of SC candidates.

\begin{figure}
\includegraphics[width=\linewidth]{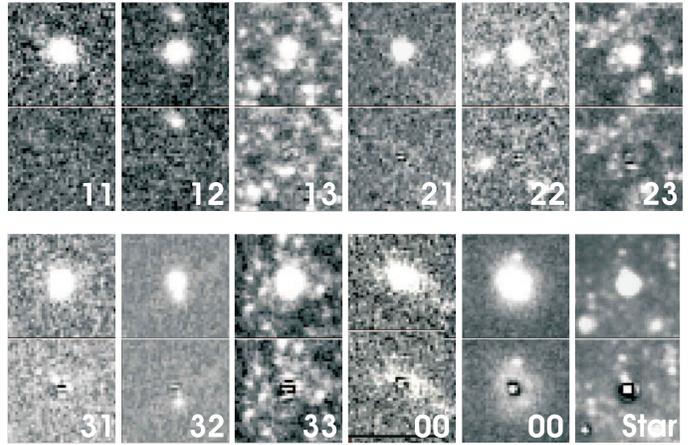}
\caption{Some examples of visual classifications codes. For each
source we show the original I-band image (upper sub-panels) and the
ishape residuals (lower sub-panels).  Objects associated to the
classes 33 (large residuals, crowded area), 100 (possible background
galaxies), and stars are rejected in the final catalogue of SC
candidates.}
\label{visual}
\end{figure}

The \verb ishape ~output parameters were used to further constrain
the list of SC candidates. In particular, we reject the objects:
\begin{itemize}
\item whose effective radii \reff differ more than 2$\sigma$ in the
two bands (Figure \ref{vi_ishape}). The $\sigma$ are derived
using standard error propagation on \verb ishape ~fitted parameters
and uncertainties. In few cases the output of the fit is very good
in both bands, so that the $rms$ is small for V and I \reff, and the
object will end up rejected by a simple 2$\sigma$ rejection
criterion. However, if the systematic uncertainties arising, for
example, from the PSF, the analytic profile chosen, etc., are taken
into account, then the two sources should be included in the
catalogue. After several tests, we decided to use a default systematic
uncertainty of $\sim$15\% \reff. Then, for the 2$\sigma$ selection we
used the maximum $\sigma$ between the $rms$ from \verb ishape , and
$\sim$15\% \reff;

\item with S/N$_{ishape}\lsim$30 \citep{larsen99};

\item with too small/large \reff. The specific properties of
Moffat profile chosen allow to estimate effective SC radii below the
$\sim$1.2 pc limit aforementioned \citep[see][and ishape user's
guide]{larsen99}.  However, we choose to adopt as lower and upper
rejection limits \reff$<$ 1.5 pc, and \reff$>$ 20 pc, respectively.
As a reference, using the data from \citet{mclaughlin05}, $\sim1$\% of
massive star clusters in the Milky Way and its satellites have
\reff$<$1.5 pc, while $\sim3$\% have \reff$>$ 20 pc;

\item elongation (major to minor axis ratio) $> 2.0$, comparable to
the maximum observed in Magellanic Clouds star clusters \citep{vdb84}.
\end{itemize}

With these selection criteria we obtained a list of $\sim$630 SC
candidates. It is worth to emphasise the strongest constraints come
from the S/N limit, and the \reff (most of the $\sim8000$ sources
are actually classified as stars, see Figure \ref{vi_ishape}).

To verify the goodness of the \verb ishape ~best fit parameters we
have visually inspected all sources, introducing a sub-classification
of sources depending on the residuals of SC candidates subtraction,
and on the crowding. The {\it visual classification} code was used for
further selections. Figure \ref{visual} shows some examples of sources
with different visual classification. The visual class code that we
used is a number composed by two digits, the first associated to the
goodness of \verb ishape ~residuals, the second to the level of
crowding. For both digits the range adopted is 1-3. For the first
digit 1/2/3 means good/satisfactory/large residuals, respectively. The
second is related to crowding with 1/2/3 meaning no crowding/less/more
than three sources within 20 pixels from the SC candidate. As an
example $visual~class=12(31)$ means good residuals and 1 or 2 sources
within 20 pixels (large residuals and no other source within 20
pixels). A default $visual~class=100$ is adopted for diffuse objects
(likely background galaxies) showing very large residuals in the
\verb ishape ~subtracted frame (see Figure \ref{visual}).
\begin{figure}
\includegraphics[width=\linewidth]{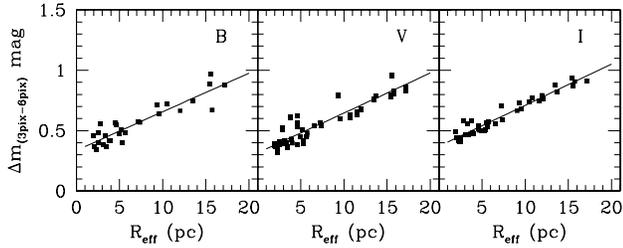}
\caption{Aperture corrections in B, V, and I-band versus the SC
candidate radius. The solid lines show the best fit linear equation.}
\label{apcorrbvi}
\end{figure}

\begin{table*}
\caption{Number of objects rejected per selection criteria}
\label{tab_sel}      
\centering                                      
\begin{tabular}{l| c c c c c c}          
\hline\hline                        
\multicolumn{7}{l}{First selection: SExtractor selection criteria}\\
\multicolumn{7}{l}{$N_{in}=17767$, $N_{out}=8081$, $N_{rej,tot}=9686$} \\
Sel. Criteria          & CLASS\_STAR=0.00 & Elong.$_{SExt.}\geq$2.5 & $m_I\geq 27.5$ &  $m_V\geq 28.5$ & $FWHM_I\geq 12$ &$FWHM_V\geq 12$  \\
$N_{rej}$              & 675              & 1652                    & 3251           &  3962           &  2452           & 3287            \\
\hline                                   
\multicolumn{7}{l}{Second selection: Ishape selection criteria}\\
\multicolumn{7}{l}{$N_{in}=8081$, $N_{out}=635$,  $N_{rej,tot}=7446$}      \\
Sel. Criteria          & 2$\sigma$\reff   & Elong.$_{ish.}>2.0$    & S/N$<$30 &   \reff$\geq$ 1.5 pc &  \reff$\leq$ 20 pc &   \\
$N_{rej}$              & 1414             & 1293                   & 5051     &  5277                & 953                &   \\
\hline                                   
\multicolumn{7}{l}{Third selection: photometric selection criteria}\\
\multicolumn{7}{l}{$N_{in}=635$, $N_{out}=277$, $N_{rej,tot}=358$} \\
Sel. Criteria          & $\Delta (V-I)> 0.15$ &  vis. class$\geq$33 & V-I $\geq 1.5$ & B-V $\geq$ 1.2 &   & \\
$N_{rej}$              & 17                   & 286                 &   66           & 19             &   & \\
\hline                                             
\end{tabular}
\end{table*}

The aperture photometry of all the sources selected has been obtained
using the Phot/Apphot task under IRAF. Due to the high, strongly
variable background in some regions of the galaxy, we made different
tests using 3, 5 and 7 pixels aperture radius, and 10, 16 pixel
annulus for background determination.  As a final choice, we adopted 3
pixels aperture for the photometry, while the background was estimated
in an annulus 16 pixels inner radius and 3 pixel width.

\begin{figure}
\includegraphics[width=\linewidth]{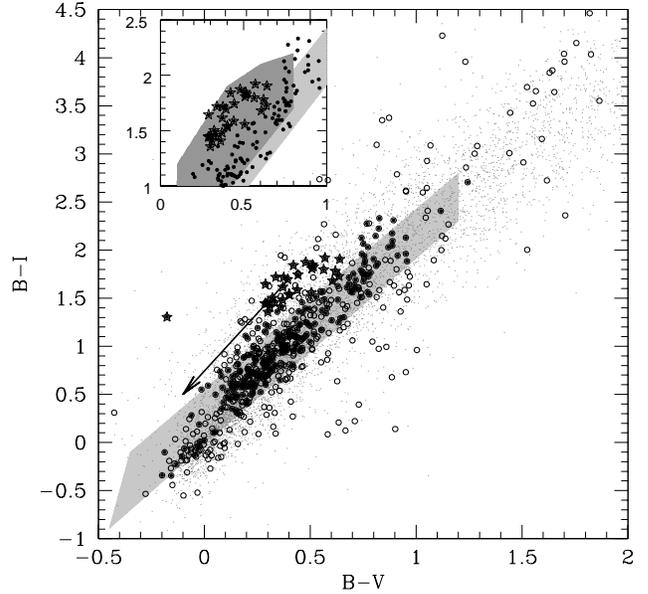}
\caption{Colour-colour diagram of the SC candidates. Dark-grey dots,
empty and full circles mark the positions of the sample after the
first, second and third (final catalogue) selections described in
Table \ref{tab_sel}. The grey shaded area marks the region of
SPoT-SSP models by \citet{raimondo05}. The arrow shows the direction of the reddening
vector. Five pointed stars are ``outlier'' SC candidates, possibly
associated to intermediate age clusters. The insert
panel shows the SC candidates (full dots), the SC outliers
(five-pointed stars), and the standard SPoT SSP models (grey-shaded
area). The dark-grey area shows the region occupied by SSP realisations
of the recent SPoT models by \citet{raimondo09}, computed paying
particular attention to the stochastic effects of TP-AGB stars. Only
ages between 0.3-2 Gyr and \feh$\sim-0.4$dex are plotted (see text for
details).}
\label{bvbi}
\end{figure}

Dealing with slightly extended objects having different radii, the
aperture correction depends on the radius of the cluster. As a
consequence, we fitted the aperture correction versus \reff. To avoid
contamination from galaxy background, uncertain \reff estimates, and
crowding, we decided to use only sources brighter than
$m_{I,3pix}=26.5$ mag, \verb ishape ~S/N$>$50, visual class $\leq 23$,
and local surface brightness background $\mu_V\geq22.5$ mag
(galactocentric radius $R_{gc}\gsim35\arcsec$). Aperture magnitude was
derived in 3 and 6 pixels ($\sim 15$ and $30$ pc) apertures over the
selected sources. Figure \ref{apcorrbvi} shows the aperture correction
derived as the difference between 3 and 6 pixels aperture magnitudes,
versus the effective radii. The best fit linear correlations shown in
the figure for each passband was then used to correct the magnitudes
of all SC candidates in B-, V-, and I-band.

The correction from 6 pixels to ``infinite'' radius, as well as the
transformations from the F435W/F555W/F814W ACS filter system to
standard B/V/I filters, and foreground extinction, were applied
following the prescriptions by \citet{sirianni05}.

Once all corrections were made, we added new constraints to further
clean the final list of SC candidates. In particular, to eliminate
objects with very high internal extinction and possible distant
background galaxies, we considered only those sources with $V-I\leq
1.5$ mag, $B-V\leq 1.2$ mag (typical maximum colours for globular
clusters), and a maximum photometric uncertainty $\Delta(V-I)\leq
0.15$ mag.  SC candidates showing large residuals and in crowded
regions, candidate background galaxies (visual class $=$33 and 100,
respectively) as well as sources located in regions with strongly
variable background (to avoid objects with photometry significantly
contaminated by internal extinction) have been rejected, too. The
final list of SC candidates matching with the required constraints is
composed by 277 objects\footnote{The complete catalogues of sources
can be retrieved upon request to the authors.}. A schematic view on
the effects of the various selection criteria applied, with more
details on the numbers, is reported in Table \ref{tab_sel}. The
table gives the number of objects before ($N_{in}$) and after
($N_{out}$) each selection criteria is applied, together with the list
of criteria adopted and the number of candidates rejected with the
chosen parameter ($N_{rej}$). It is worth noting that the same object
might satisfy two or more rejections, thus the total number of objects
rejected, $N_{rej,tot}$, is not simply obtained by summing up the
numbers in the $N_{rej}$ lines. The positional, photometric and
structural properties of the final catalogue of sources are reported
in Table \ref{tab_final}.

\section{Discussion}

\begin{figure}
\includegraphics[width=\linewidth]{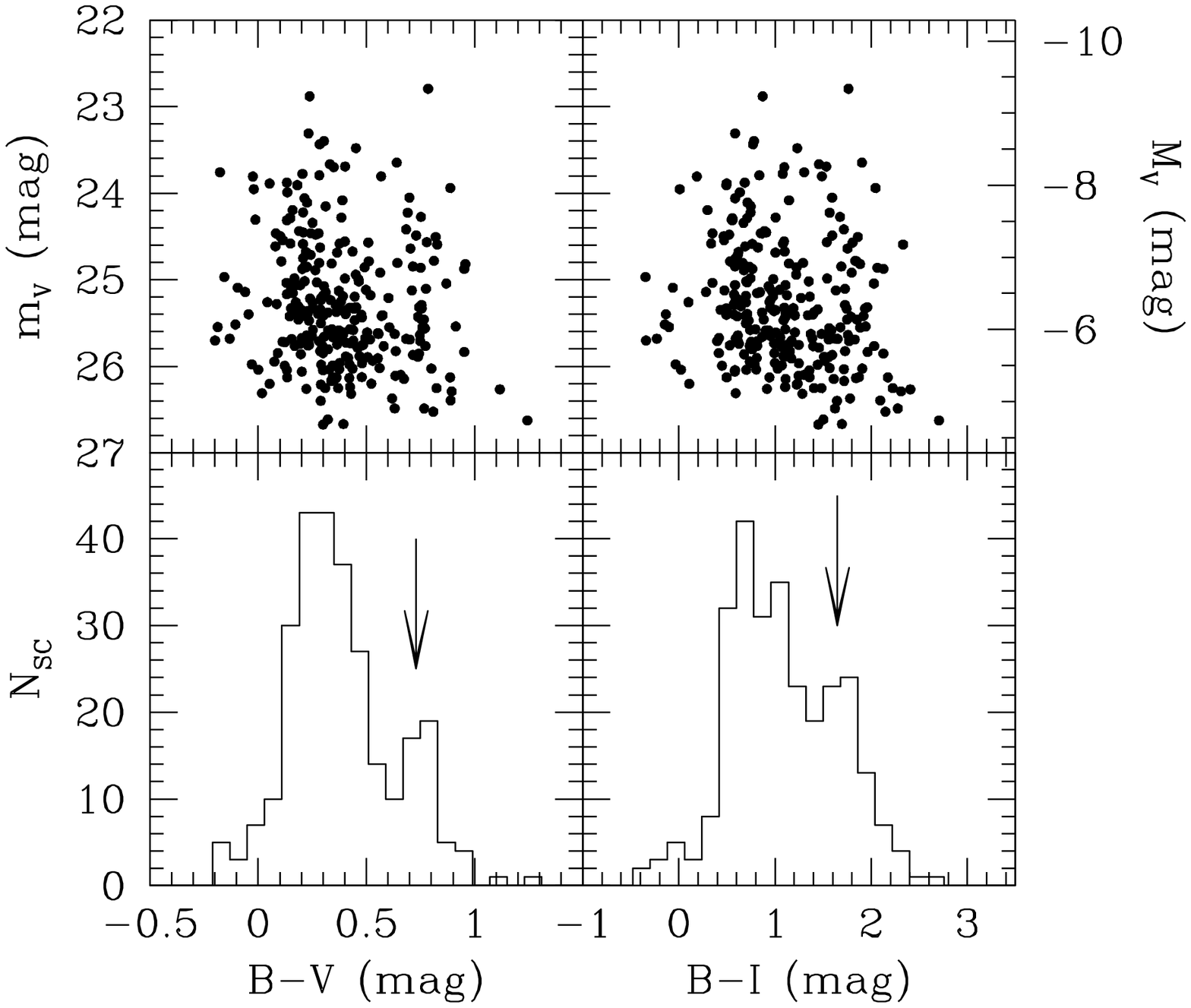}
\caption{Left panels: B-V colour magnitude diagram (upper panel) and
colour histograms (lower panel) of SC candidates. Right panels: as
left, but for B-I.  The arrows indicate the median colour of GGCs.}
\label{col_mag}
\end{figure}

\subsection{Analysis of Colours}

\begin{table*}
\center
\caption{Structural and photometric properties of selected SC candidates. {\it[See the electroniv version of the Journal for the complete version of the Table.]}}
\begin{tabular}{r c c r r c c c c c c c c}
\hline \hline
ID &    RA (deg) & Dec (deg)  & $R_{gc}(\arcsec)$& \reff(pc) &$\pm$& $m_{V}$ & $\pm$ & B-V & $\pm$ & V-I & $\pm$ & v.c. \\
\hline
5 &   161.7465 &  17.2766 &      71.65 &     2.04 &  0.25 &    25.39 &  0.01 &     0.23 &  0.02 &     0.28 &  0.01 &    21 \\
10 &   161.7527 &  17.2907 &      79.03 &     5.50 &  0.45 &    26.15 &  0.02 &     0.67 &  0.06 &     0.94 &  0.03 &    12 \\
13 &   161.7513 &  17.2858 &      69.83 &     4.68 &  0.92 &    25.40 &  0.02 &    -0.04 &  0.03 &    -0.09 &  0.05 &    23 \\
14 &   161.7552 &  17.2964 &      91.56 &     5.58 &  0.41 &    25.98 &  0.01 &    -0.03 &  0.03 &     0.00 &  0.03 &    22 \\
15 &   161.7556 &  17.2980 &      96.25 &    13.67 &  0.21 &    24.57 &  0.01 &     0.51 &  0.01 &     1.30 &  0.01 &    21 \\
17 &   161.7537 &  17.2915 &      79.24 &     6.86 &  0.45 &    25.77 &  0.01 &     0.58 &  0.04 &     0.82 &  0.02 &    23 \\
18 &   161.7527 &  17.2879 &      71.40 &     6.62 &  0.18 &    24.46 &  0.01 &     0.28 &  0.01 &     0.58 &  0.01 &    22 \\
21 &   161.7506 &  17.2792 &      60.20 &     6.35 &  0.44 &    26.26 &  0.02 &     0.22 &  0.04 &     0.69 &  0.03 &    12 \\
22 &   161.7523 &  17.2850 &      65.42 &     2.13 &  0.38 &    25.09 &  0.01 &    -0.09 &  0.02 &     0.03 &  0.01 &    22 \\
23 &   161.7545 &  17.2898 &      72.80 &     4.50 &  0.34 &    25.69 &  0.02 &     0.70 &  0.06 &     0.86 &  0.03 &    23 \\
24 &   161.7519 &  17.2828 &      61.81 &     4.00 &  0.45 &    25.76 &  0.02 &     0.17 &  0.04 &     0.56 &  0.02 &    23 \\
25 &   161.7530 &  17.2856 &      64.93 &     3.84 &  0.40 &    25.58 &  0.01 &     0.38 &  0.03 &     0.56 &  0.02 &    12 \\
29 &   161.7495 &  17.2750 &      60.90 &     3.89 &  0.46 &    26.04 &  0.01 &     0.30 &  0.03 &     0.40 &  0.02 &    22 \\
30 &   161.7550 &  17.2895 &      70.74 &    13.23 &  0.46 &    26.17 &  0.02 &     0.34 &  0.06 &     1.38 &  0.03 &    12 \\
33 &   161.7529 &  17.2837 &      60.83 &     3.06 &  0.10 &    25.40 &  0.01 &     0.18 &  0.03 &     0.39 &  0.02 &    23 \\
34 &   161.7573 &  17.2952 &      84.98 &     2.54 &  0.08 &    24.64 &  0.01 &     0.70 &  0.02 &     1.05 &  0.01 &    22 \\
36 &   161.7603 &  17.3021 &     105.49 &     2.74 &  0.57 &    25.90 &  0.01 &     0.42 &  0.02 &     0.39 &  0.01 &    22 \\
37 &   161.7549 &  17.2870 &      64.02 &     5.79 &  0.24 &    26.37 &  0.02 &     0.62 &  0.07 &     1.16 &  0.03 &    13 \\
\hline
\label{tab_final}
\end{tabular}
\end{table*}

\begin{figure}
\includegraphics[width=\linewidth]{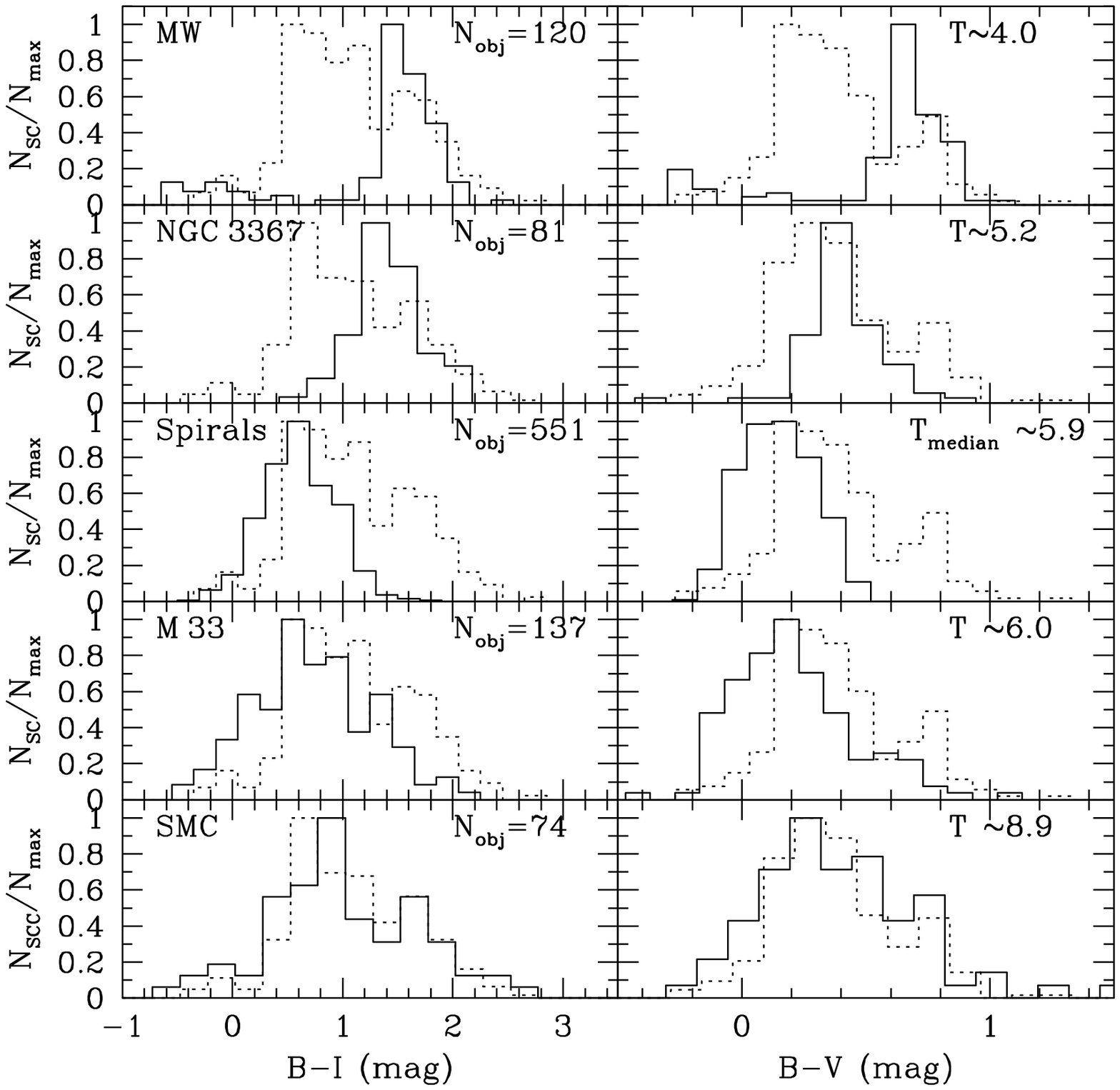}
\caption{Colour histograms of the comparison galaxies (as labelled in
left panels). Dotted lines show \object{NGC 3370} SC candidates. For \object{NGC 3367}
and \object{SMC} a different binwidth is used, because of the smaller number of
SCs (see $N_{obj}$ labels in each panel).}
\label{col_hist}
\end{figure}

Figure \ref{bvbi} shows the colour-colour diagram of the final
catalogue of SC candidates. In the figure it is shown also the effect
of the various steps of the selection. Dark-grey dots, empty and
filled circles mark the position of the sample after the selections
based on SExtractor, \verb ishape ~, and photometric parameters,
respectively. The full circles in the figure refer to the final
selected sample of SC candidates. Also shown in the figure, with a
grey area, is the region of the SPoT Simple Stellar Populations models
for metallicity between \feh$\sim-2.5 \div 0.3$ dex, and ages in the
interval 25 Myr $\div$ 14 Gyr \citep[see][for more details on SPoT SSP
models]{raimondo05}. As expected, the data overlap to SSP models,
since SC are, at least in first approximation, well represented by
SSPs. It must be emphasised, however, that such overlap is not a
trivial result. In fact, in the selections described in the previous
section, the only constraint on colour was on the maximum $V-I$ and
$B-V$ allowed.

One interesting feature appearing in Figure \ref{bvbi} is the
non-negligible number of objects located at $B-V\sim$0.4 and
$B-I\sim$1.6 (five pointed stars in the figure). After different
counter-checks, we concluded that there is no obvious reason to reject
these clusters from the final catalogue. As shown by the arrow in the
panels, even an internal extinction E(B-V)$\sim$ 0.5 mag would not
shift these SC candidates in the region of SSP models. Moreover, such
high extinction is unlikely since such objects do not appear obviously
associated to regions of high internal extinction, while they are
rather evenly distributed over the area of the galaxy. One possible
interpretation of these objects is that they are intermediate age SCs
($0.3 \leq t (Gyr) \leq 2$). Within this age interval, the emission of
Thermally Pulsating AGB stars (TP-AGBs) dominates the light emitted by
the SC, in particular in the long-wavelength regime \citep[$\gsim8000
\AA$; see, e.g.][and references therein]{maraston05}.  However, the
presence of stars in such evolutionary stage is strongly influenced by
stochastic effects. As a consequence, the stochasticity due to the
number of cool TP-AGBs can significantly affect the integrated colours
of the stellar system. To highlight such behaviour, the insert panel
in Figure \ref{bvbi} shows the sample of selected SC candidates, and
the ``outliers'' at $B-V\sim$0.4 and $B-I\sim$ 1.6 mag (five-pointed
stars), together with the region where the recent SSP models by
\citet{raimondo09} for intermediate ages and \feh$\sim-0.4$ dex are
distributed (dark-grey area). The new set of models used has been
computed with an updated version of the SPoT code, paying particular
attention to the statistical effects generated by Thermally Pulsating
AGBs. In the figure, the dark-grey area shows the distribution of
models when the stochastic effects due to TP-AGB stars are properly
taken into account. As shown in the insert, these models predict the
existence of the outliers \citep[optical models have been provided as
a private communication by G. Raimondo, for more details on models and
near-IR colours see][]{raimondo09}. In a following section we will
discuss more in details the data to models comparison for the whole
sample of SC candidates.

Figure \ref{col_mag} shows the colour magnitude diagrams and colour
histograms for the SC candidates. It is not surprising that the whole
population of SC candidates in \object{NGC 3370} does not have a unimodal
colour distribution, showing the presence of sub-peaks possibly
associated to different SC sub-populations.

As a matter of fact, since \object{NGC 3370} is a late type Sc spiral, it is
expected that this class of galaxies is actively forming stars, and
star clusters. Thus, populations of young/intermediate age SC are
expected to host in the galaxy, in addition to the old population of
Globular Clusters (GCs). To highlight this point, in the colour
histograms shown in Figure \ref{col_mag}, we have added an arrow
showing the median colours of Galactic GCs \citep[GGCs,
$(B-V)_0=0.73\pm0.12$, $(B-I)_0=1.65\pm0.12$ mag for $\sim$90 GGCs
with $M_V\lsim-5.0$ mag, data from ][]{harris96}. It is rather
obvious, then, to conclude that the red peak observed in the lower
panels is actually due to the GC population in the galaxy.

If the intermediate peak and the blue tail of SC candidates are taken
into account, still using as reference the system of young open
clusters (OCs) in the Galaxy, we find that the OCs at
$B-V\sim$0.35, and $B-I\sim$0.85 mag have ages $\log t(yr)\sim
8.4$ (based on only 3 Galactic OCs), while blue OCs at
$B-V\sim$-0.15, and $B-I\sim$-0.25 mag have on average $\log
t(yr)\sim 7.0$ (based on 11 OCs)\footnote{Ages and colours for OCs are
from the WEBDA database, available at http://www.univie.ac.at/webda}.
Given the results of such comparison with Galactic star clusters, it
is reasonable to conclude that the colour peaks observed in Figure
\ref{col_mag} are mainly driven by age differences between the
blue/intermediate/red SC candidates.

\begin{figure}
\includegraphics[width=\linewidth]{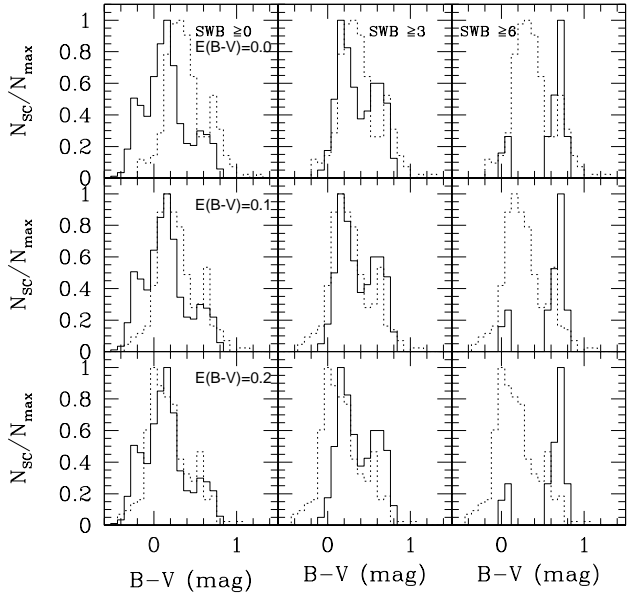}
\caption{B-V colour histograms of \object{LMC} Star Clusters (solid lines)
and \object{NGC 3370} (dotted lines). The vertical blocks refer to \object{LMC}
clusters selected according to the SWB class: the first, second and
third row refer to \object{LMC} clusters with SWB$\geq 0, 3, 6$, respectively
(see labels in upper panels). Horizontal blocks assume different
average internal extinctions for \object{NGC 3370} SC candidates: first, second and
third row have $\langle E(B-V)_{internal}\rangle=0.0, 0.1$, and 0.2
mag, respectively (see label in leftmost panels).}
\label{hist_lmc}
\end{figure}

The situation, however, might be complicated by the possible presence
of residual internal extinction. Since the galaxy is oriented nearly
face on, the effect of internal extinction is minimised. Moreover, as
discussed in previous section, to reduce possible extinction present
in the dusty, star forming regions we rejected the sources located in
areas of strongly varying background, or where dusty patches can be
recognised. The presence of the colour peaks discussed above is an
additional indirect probe against a possible strong contribute of
internal reddening: a significant amount of dust extinction, in fact,
should smear out such features. However, it is not possible to exclude
at all the presence of some residual internal extinction.  To have
other indications about the possible nature of the colour range and
sub-peaks in \object{NGC 3370}, we compare the $B-I$ and $B-V$ colour
histograms of SC candidates with the same data for other galaxies,
some of which are corrected for internal reddening. In particular, we
use the data from $1)$ Milky Way's GCs \citep{harris96}, and Open
Clusters \citep{lata02}; $2)$ \object{NGC 3367} from \citet{garciab07}; $3)$
the data of $\sim$ 20 Spirals from \citet{larsen99}; $4)$ \object{M 33} data
corrected for reddening \citep{park07}; $5)$ \object{SMC} Star Clusters from
\citet{rafelski05}; $6)$ \object{LMC} data from \citet{bica96}. All the
data used are in (or are transformed to) the standard photometric BVI
bands. The comparisons between different galaxies are shown in Figure
\ref{col_hist} and \ref{hist_lmc}. In order to compare only SCs in the
luminosity range of the SC candidates considered here, we have
excluded all SCs having total absolute magnitude $M_V \gsim -5$ mag.

Figure \ref{col_hist} shows $B-I$ and $B-V$ colour histograms of the
SC in the \object{MW}, \object{NGC 3367}, the spirals from
\citet{larsen99}, \object{M 33} and the \object{SMC}. The galaxies
are ordered according to their morphological T-Type, all histograms
are normalised to have 1 in the most populated bin. In each panel of
the figure the colour histograms of SC candidates in
\object{NGC 3370} are shown with dotted lines.

\begin{figure}
\includegraphics[width=\linewidth]{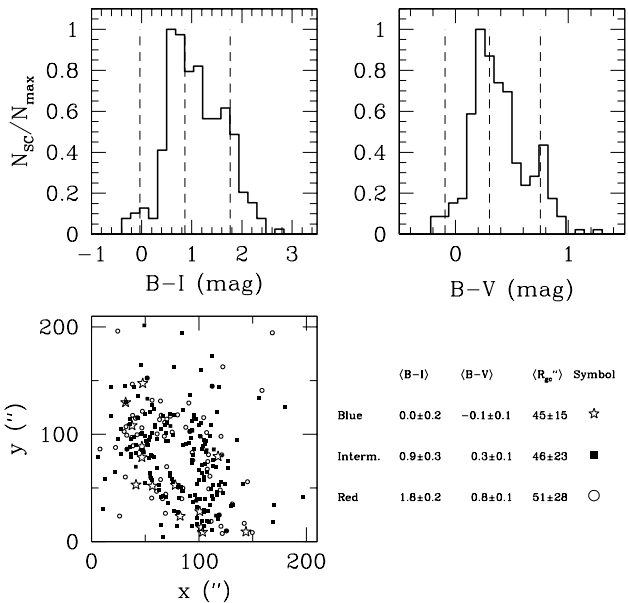}
\caption{Upper panels: colour histograms of SC candidates, the
position of the three relative peaks is emphasised in each panel with
dashed lines. Lower-left panel: spatial distribution SC
candidates selected according to their membership to the colour
sub-peaks (see text). Five pointed stars/full squares/circles mark
candidates in the blue/intermediate/red tail respectively.}
\label{col_site}
\end{figure}

By inspecting the colour histograms it is easy to reveal similarities
and differences between the system of SC candidates in
\object{NGC 3370} and in other galaxies. One example is represented
by the red sub-peak, which appears quite similar to the red sub-peak
in the colour distribution of \object{SMC} clusters. Using the age
estimates by \citet{rafelski05} we find that the $\sim20$ SMC clusters
with $B-I\gsim 1.6$ mag have a median age $t\sim 5$ Gyr. The range of
colours generally shown by other systems is quite similar to that of
\object{NGC 3370}. The only exception is represented by the colour
histograms of the SCs from \citet{larsen99}. However, the selection
criteria used by \citeauthor{larsen99} exclude SCs with $B-V\geq
0.45$, which represents $\sim$45\% of our sample of objects. In
addition, the very blue/young stellar clusters are rejected by Larsen,
adopting a lower limit to the $H_{\alpha}-R$ colour. Thus, we are lead
to conclude that, rather than a systematic difference between our data
and \citeauthor{larsen99}'s, the observed differences are due to the
intrinsically different SCs selection.  In any case, the peak of the
distributions of \citeauthor{larsen99}'s data appears similar to the
main peak of \object{NGC 3370} SCs, though slightly shifted towards
bluer colours.

The SCs data of \object{NGC 3367} - the galaxy of the sample
morphologically more similar to \object{NGC 3370} - do not show
obvious multi-modal colour distributions. The range of colours, and
the position of the main peak are roughly similar to
\object{NGC 3370}. As for what concerns \object{SMC} clusters, the
colour distributions span over the same range, and the peak of SC
sub-systems seems to be recognised here as in \object{NGC 3370}.

\begin{figure}
\includegraphics[width=\linewidth]{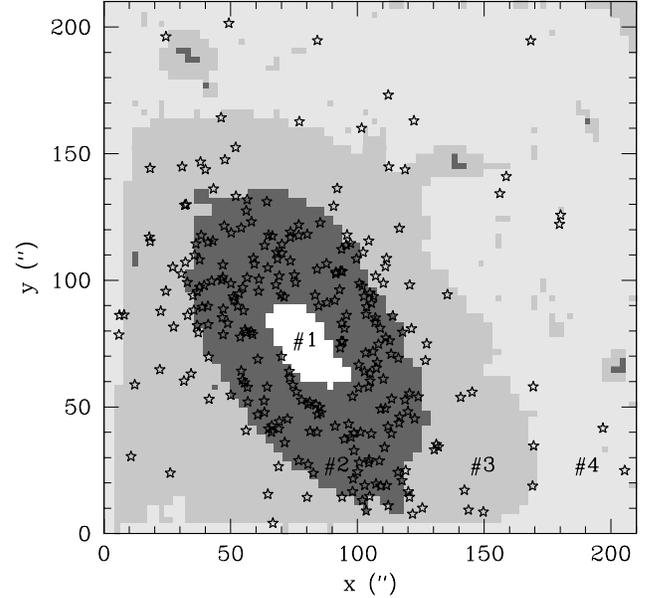}
\caption{Annuli division used to compute completeness functions. The
regions are characterised by their median surface brightness:
$\mu^{\#1}_I<$20, $\mu^{\#2}_I$ between 20$\div$23, $\mu^{\#3}_I$
between 23$\div$26.5 and $\mu^{\#4}_I>26.5 mag/arcsec^2$. Each annulus
is labelled and identified with a different shade of grey. Also the
loci of SC candidates are shown with five pointed stars. No SC
candidate lies in the innermost annulus (\#1, white area).}
\label{positions}
\end{figure}

We provide in Figure \ref{hist_lmc} a separate comparison to \object{LMC}
data. The reason for keeping \object{LMC} data aside is that, in this case, we
can play with the age of the selected clusters by using the SWB class
\citep{searle80} of \object{LMC} clusters, and also make a test on the internal
extinction in \object{NGC 3370}. The figure shows the $B-V$ colour histograms
of \object{LMC} clusters, corrected for internal extinction, and \object{NGC 3370} SC
candidates. Each row of panels in the figure shows \object{NGC 3370} cluster
candidates dereddened adopting a different average internal
extinction, $\langle E(B-V)_{internal} \rangle = 0.0,~0.1$ and 0.2 mag
from upper to lower panel, respectively.  In addition, assuming the
SWB classes from \citet{bica96}, each column of panels is obtained
assuming a lower limit to the SWB. Specifically, the first column
includes all SWB classes, while the other two include only \object{LMC}
clusters with SWB$\geq$3 and 6, respectively (labels in the uppermost
panels).  All panels show \object{NGC 3370} colour histograms (dotted lines)
shifted according to the internal reddening assumed, and the \object{LMC}
clusters colour histograms using the SWB selection criteria (solid
lines). Some hints on age distributions and, possibly, internal
extinction of the SC candidates can be derived by a careful inspection
of this figure. The upper panels show that if old \object{LMC} clusters are
taken into account (SWB$\geq6$, t$\gsim 10^9$ yr), there is a good
match between \object{NGC 3370} red peak and the old clusters in \object{LMC}. Thus, we
have an additional indication that this population is, in fact,
composed by old GC candidates.  It is worth to notice, in the case of
$\langle E(B-V)_{internal} \rangle \sim 0.1$ and SWB$\geq 3$ (central
panel in the figure), the similarity between the main colour peaks of
the two galaxies.  Clearly, because of the differences between the two
galaxies, it is not possible to conclude whether or not such good
matching is a consequence of an $average$ internal extinction of
$\sim$0.1 mag on a SC system in \object{NGC 3370} with an age distribution
similar to \object{LMC}.  It is, however, reasonable to take this value as an
upper limit for $\langle E(B-V)_{internal} \rangle$. Furthermore, the
lower panels, which assume larger internal extinction, show that such
average reddening would move the main peak to a colour corresponding
to ages of the order of $\log t(yr)\sim 7.5$ - obtained taking MW OCs
and \object{LMC} clusters ages as reference. Although young SCs are expected to
be present in late type, actively star forming galaxies, it is not
reasonable to conclude such predominance of very young objects. The
position of the red peak is not an issue in the latter case, since it
is expected that old clusters are, on average, less affected by dust
extinction being far from actively star-forming regions. Thus, the
shift of the red peak for $\langle E(B-V)_{internal} \rangle=0.2$ mag
is not probable.  However, since this test is carried out using a {\it
common} average value for internal extinction, we cannot rule out high
extinction values on {\it single} bluer/younger cluster candidates.

In conclusion, the comparison of the colour histograms for \object{NGC 3370}
SC candidates with other SCs from literature shows that the range of
colours appear normal with respect to similar class of
galaxies. Clearly, the comparisons shown cannot be used to constrain
ages, or any residual internal reddening, because of the intrinsic
differences existing between the galaxies. However, we can give some
limits, in particular, it is reasonable that \object{NGC 3370} hosts three
sub-populations of SC candidates: one with colours typical of old GCs,
one with very blue colours, characterising a population of very young
SCs, and a rich population of intermediate age SC candidates. This
expectation is partly supported by the spatial distribution of the SC
candidates in the three colour peaks, shown in Figure
\ref{col_site}. In the figure we used the position of the three
relative maxims to define a preliminary centre of each sub-peak in B-I
and B-V. Then, the median and standard deviation of the sole SCs
within half-distance from the sub-peaks is obtained (dashed-lines in
the upper panels). All SC candidates within
$2\sigma$ from the median value of the peak in both colours are then
selected. The SCs selected in this way are shown in the lower-left
panel of the figure: five pointed stars/full squares/circles mark
candidates in the blue/intermediate/red tail respectively. By
inspecting the spatial distribution of clusters, we recognise that the
SC candidates in the blue tail are mostly located near the galactic
centre, while the star clusters in the other colour bins are more
spread over the galaxy, with the ones in the red peak being on average
more distant from the centre of the galaxy, as expected for GC
candidates. The median effective galactocentric distance of each SC
class is estimated and reported in the figure.

\begin{figure}
\includegraphics[width=\linewidth]{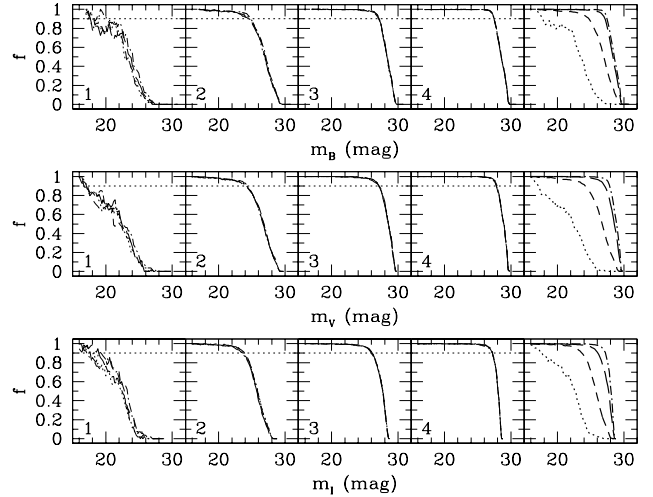}
\caption{B, V and I-band completeness functions for the each of the
four annuli considered (labelled in the panels). For each one of the 4
annuli, we have carried out several numerical experiments shifting the
grid of artificial stars. Different experiments are shown with
different line types in each panel. The horizontal dotted line marks
the 90\% completeness limit in the annulus. The rightmost panels show
the median completeness function for each annulus and filter.}
\label{cfunctions}
\end{figure}

\subsection{Luminosity Functions}
\label{sect:lf}

\begin{figure*}
\includegraphics[width=6cm]{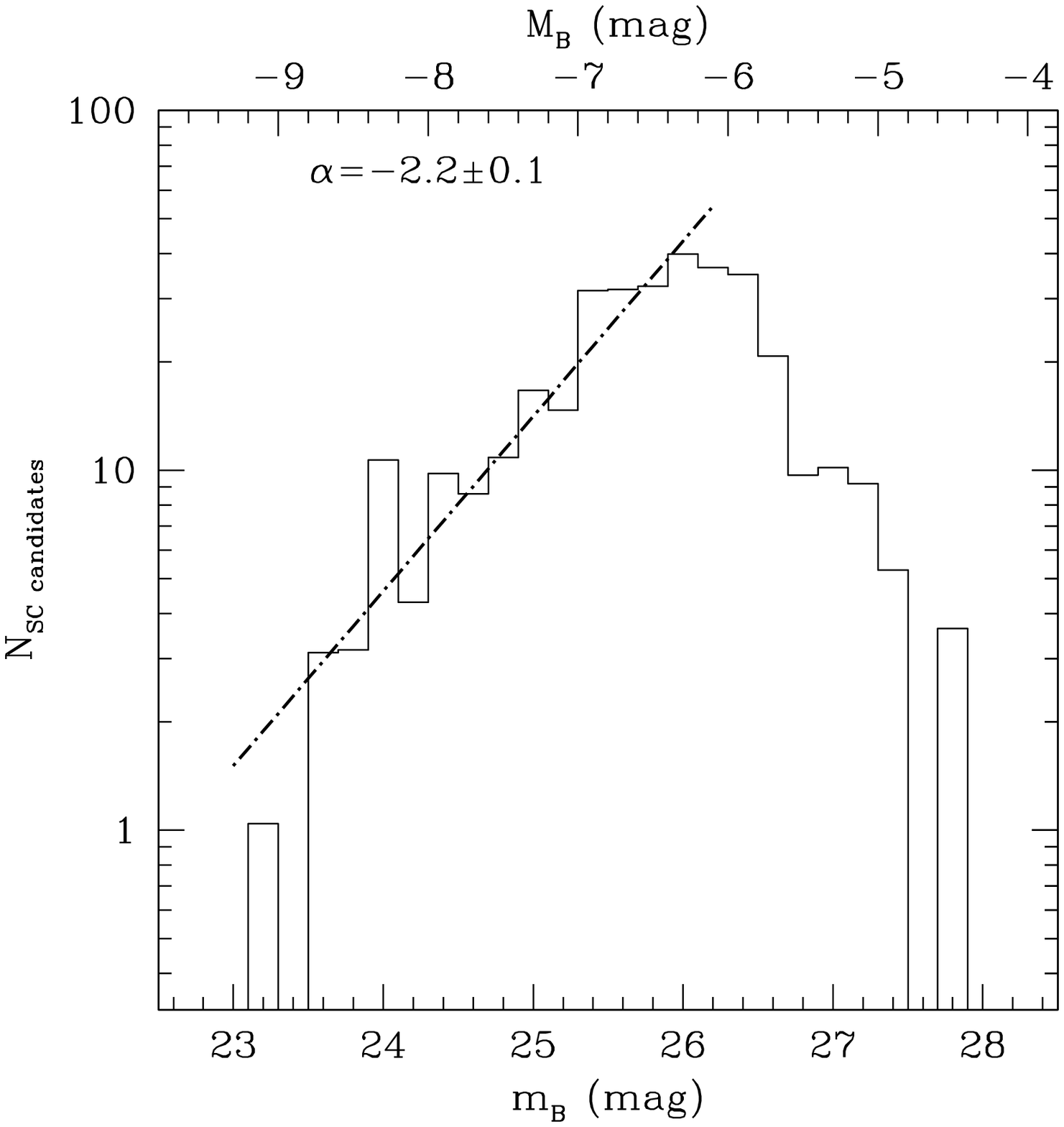}
\includegraphics[width=6cm]{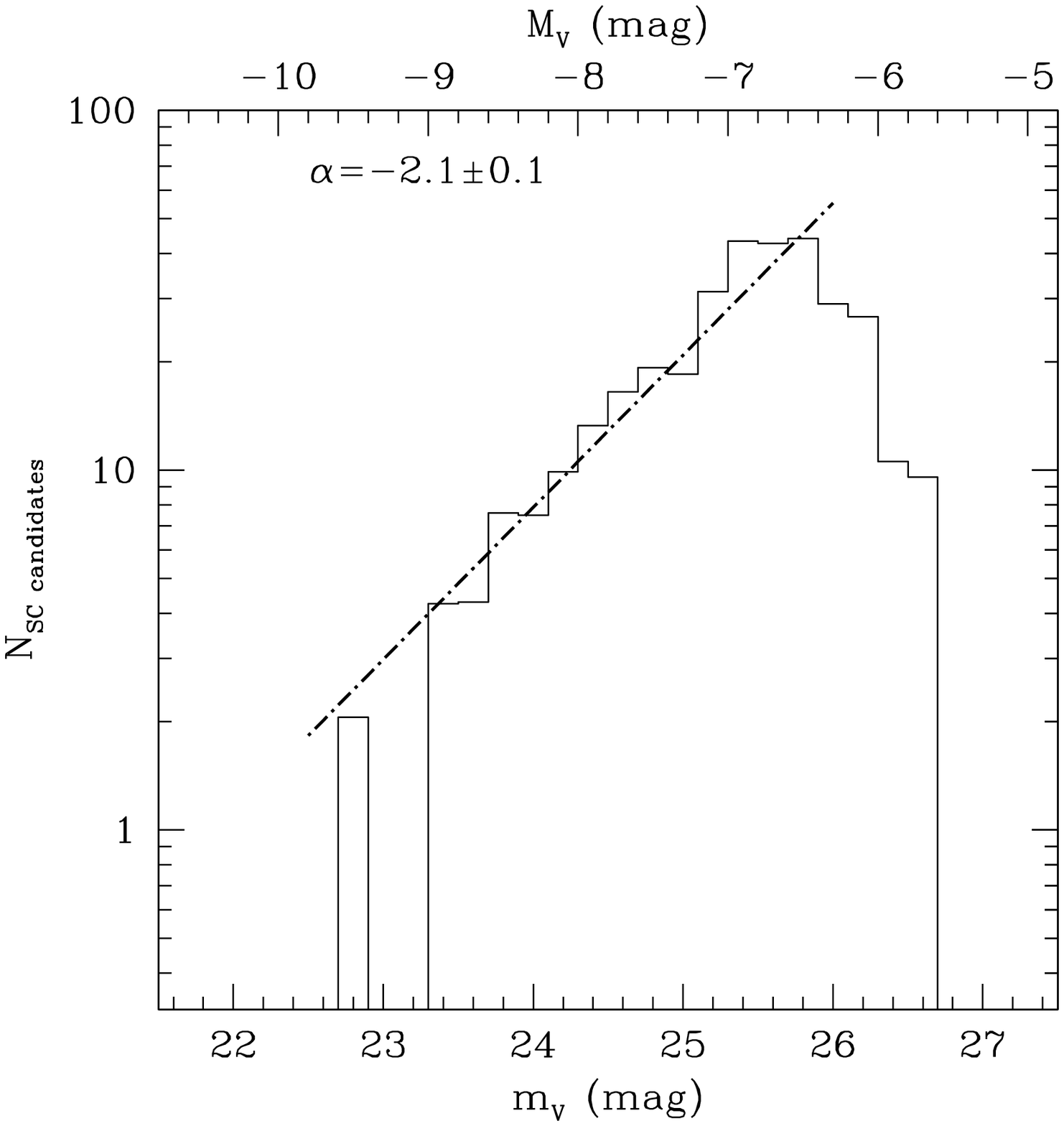}
\includegraphics[width=6cm]{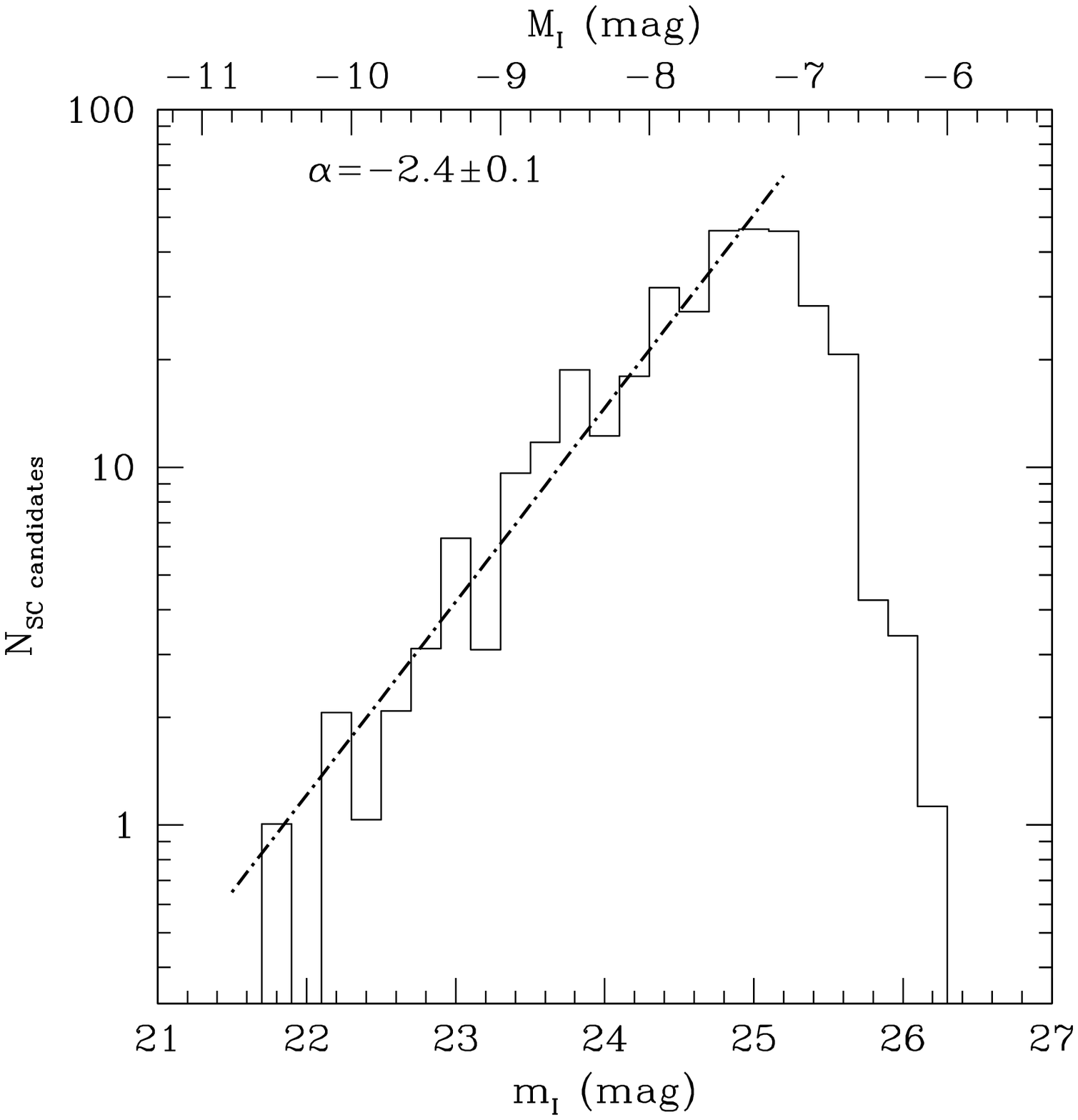}
\caption{Left panel: B-band Luminosity Function of SC
candidates. The dot-dashed line shows the power law fit to the data. The
exponent of the fit is also reported in the panel. Middle panel: as
upper panel, but for V-band data. Right panel: as upper panel, but for
I-band data.}
\label{lg_pl}
\end{figure*}

Starting from the work by \citet{elson85} on \object{LMC} SCs, empirical data
from a large sample of spiral and irregular galaxies have shown that
the Star Cluster Luminosity Function (SCLF) usually follows a power
law, with a sharp turnover at faint magnitudes. In this section we
analyse the LF profile of the SC candidate system in \object{NGC 3370}.

To properly study the SCLF, we need to take into account the
completeness corrections in the different bands. However, because of
the selection criteria adopted, this is a complex task. In our case,
in fact, the completeness functions should take into account $i)$ the
photometric completeness, $ii)$ the selection on sizes, and $iii)$ the
visual identification. While the first two contributions to the
completeness can be somehow obtained from numerical experiments, the
third one is less straightforward to derive, if feasible at all.
Moreover, the completeness fraction is different for objects of
different sizes.  For such reasons we choose to adopt the following
approach: the galaxy is divided in four concentric regions, selected
on the basis of the median surface brightness of the galaxy. In
particular, we choose $\mu^{\#1}_I <$ 20, $20 < \mu^{\#2}_I \leq 23$,
$23 < \mu^{\#3}_I \leq 26.5$, and $\mu^{\#4}_I > 26.5~mag/arcsec^2$ as
separation between the regions \#1, \#2, \#3, and \#4. Figure
\ref{positions} shows the four regions and the position of SC
candidates. In each of the four annuli, then, we run completeness
tests separately. To add artificial SCs we used DAOPHOT/ADDSTAR,
adopting a template SC candidate derived from the best candidates in
our list. Accurate completeness functions would need different
numerical experiments obtained with SC templates of various \reff. We
have not carried out such detailed analysis. It is worth to remark,
though, that the median size of template SC, $\langle R_{eff}
\rangle\sim 3$ pc, is representative of the vast majority of the SC
candidate population (see next section).

To derive completeness functions, SC candidates at fixed magnitudes
are artificially added in the frame on a grid of 100 pixel separation
both in x and y coordinates, the photometry of artificial sources is
then obtained in the same way as for the original data. The comparison
of the input and output photometric catalogues is used to evaluate the
completeness fraction $f\equiv N_{SC~detected}/N_{SC~added}$. The
experiments are repeated for $i$) different magnitudes (between 15
and 30 mag, with steps of 0.1 mag), and $ii$) for different positions
of the grid of stars. The results of the completeness tests are shown
in Figure \ref{cfunctions}. The completeness fraction is quite low in
the innermost annulus (\#1), though we recall that no source in the
final catalogue lies in such area of high, and highly variable galaxy
background (Figure \ref{positions}).

Using the completeness functions so derived, we estimated the
corrected LF of SC candidates, shown in Figure \ref{lg_pl}.  The
observed sharp falloff is an artifact due to the sample incompleteness
at faint magnitudes. In the three panels of the figure we plot a fit
to the data, the slope $\alpha$ obtained from a power law fit
$dN/dL\propto L^{\alpha}$ is also reported. To minimise the effect of
incompleteness, the upper limit of the fit is chosen so that the
completeness fraction in the innermost useful annulus, \#2, is
$f\gsim$0.5 \footnote{The average $f$ between annuli \#2, \#3, and \#4
at the faint limit is larger than 0.8 in all bands.}. Notwithstanding
the complex issue of completeness, the power law fit to the data
provides exponents $\alpha\sim-2.2$ similar to what typically found in
spirals and starburst galaxies
\citep[e.g.][]{miller97,whitmore99,larsen02,degrijs03}.  We also
find, for all three passbands considered, steeper slopes on the bright
sides of the LFs, with power law exponents as large as
$\alpha\sim3.0$. In all cases, however, the new exponent agrees within
1$\sigma$ uncertainty with the single power-law slopes reported in the
figure. The presence of a bend in the SCLF has already been found in
various galaxies. As discussed in details by
\citet{gieles06a,gieles06b}, a SC system originating from a truncated
mass function (MF), possibly caused by a truncated MF in the giant
molecular clouds, is expected to show such a bend, with steeper slopes
on bright side of the LF. One other prediction of a truncated MF is
the steepening of the LF at longer wavelengths. In spite of the
uncertainties on the fitted slopes, this behaviour seems to be found
in our data. Nevertheless, comparing maximum cluster luminosity
($M_{V,max}\sim-9.5$ mag) with the number of clusters brighter than
$M_V\sim-8.5$ mag ($N(M_V<-8.5)=13$), as proposed by \citet[][see also
the references therein]{gieles06a}, we find that the luminosity of the
brighter cluster in \object{NGC 3370} can be accounted for by a size-of-sample
effect, with a single power-law distribution \citep[see Fig. 16
in][]{gieles06a}.

The consistency of the SCLF with a power law indicates no preferred
scale for cluster luminosity (i.e. mass). However, ``passive''
evolutionary processes like tidal interactions, dynamical friction,
and the ``active'' evolution of stars in the cluster are expected to
alter the shape of the SCLF. On the contrary, the luminosity function
of GCs (GCLF) in both elliptical and spiral galaxies shows a nearly
universal turn-over at $M_V\sim -7.4$ mag \citep{harris01,richtler03},
implying that present GCs have a preferred mass scale. By
inspecting the data in Figure \ref{lg_pl}, we do not see any
peculiarity near the regions of the GCLF peaks ($m_V\sim25$ mag, and
$m_I\sim24$ mag), although this might well be a consequence of the
poor number of GC candidates in our final catalogue
(Sect. \ref{sect:gc}) and the dominance of young objects.

\begin{figure}
\includegraphics[width=\linewidth]{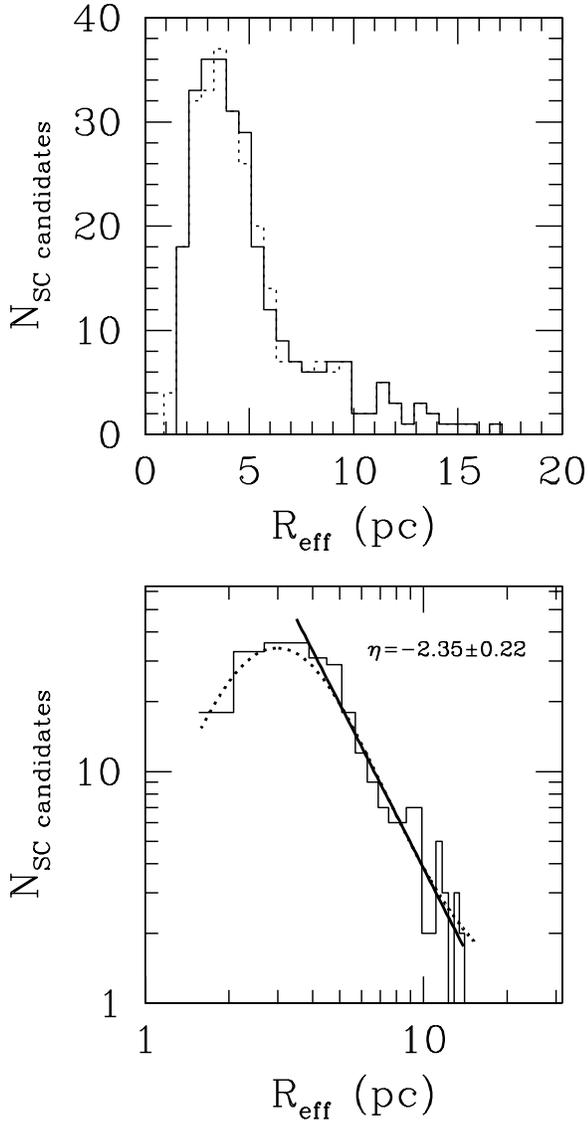}
\caption{Upper panel: \reff distribution for SC candidates. Solid line
histogram shows the mean \reff obtained from I and V band data, the
dashed-line shows the histogram obtained coupling B-, V- and I-band
data. Lower panel: As upper panel but in $\log$ scales. The thick
lines show a power law (only for \reff$\gsim 4$ pc, solid line), and a
log-normal fit to the data (all \reff, dotted line).}
\label{rad_hist}
\end{figure}

\begin{figure}
\includegraphics[width=\linewidth]{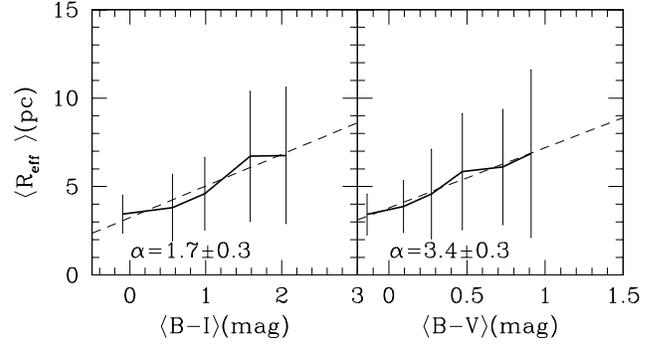}
\caption{Median \reff estimated per colour bins. The slope of the
linear fit to data and its $rms$ are reported in the panels.}
\label{reffcol}
\end{figure}

One more piece of information comes from the study of the
correlation between the SCLF and Star Formation Rate (SFR) of the
galaxy. As shown by different authors
\citep[e.g.][]{larsen00,billett02,larsen02}, some properties of the SC
system are correlated to the SFR, with the efficiency of cluster
formation being larger in galaxies with higher SFR. In particular,
\citet{larsen02} analysed the maximum luminosity expected from a
random sampling of the LF (assuming a power law exponent
$\alpha=-2.4$) versus the SFR per unit area. The comparison of
predictions to observations showed a satisfactory agreement. Using the
LF shown in Figure \ref{lg_pl} we find that the maximum observed SC
candidate magnitude is $M_V^{max}\sim-9.5$ mag. Assuming the SFR=0.08
$M_{\sun}~yr^{-1}$ by \citet[][a {\it ``crude calculation''} based on
the K-band properties of bright knots in the galaxy]{grosbol08}, and
an area A$\sim240~kpc^2$, we find that \object{NGC 3370} data fall in the same
observational regions as the spirals in the Figure 16 of
\citet{larsen02}, without any noticeable scatter with respect to other
observational data. Moreover, the behaviour with respect to the
predictions presented in \citet{larsen02}, based on sampling
statistics, is offset by nearly the same amount ($\sim$0.9 mag, or
$\sim$0.7 mag, depending on the model) as other spirals.

\subsection{About effective radii and spatial distribution}
In this section we analyse the \reff versus various other
characteristics of the SC candidates, or the host environment.  Let us
first inspect the \reff distribution of the system. 

Upper panel of Figure \ref{rad_hist} shows the distribution of \reff
using both the radii obtained from the weighted average of V and I
values (solid histogram), and those obtained using all BVI data
(dotted line). The differences between the two distributions is
negligible, mainly because the B-band \reff have larger uncertainties,
thus lower weights. As anticipated, we adopted the V and I averaged
\reff values.

The lower panel of Figure \ref{rad_hist} shows the distribution of
\reff in logarithmic scale. A power law fit to the data (solid line)
for \reff larger than $\sim$3 pc, provides an exponent
$\eta=-2.35\pm0.22$. Such value is consistent with similar results for
other galaxies. For example, \citet{bastian05} found $\eta=-2.2\pm0.2$
for the star clusters in \object{M 51}, although they fit with the same power
law the whole set of \reff ($2~pc \leq R_{eff} \leq 15~pc$). In Figure
\ref{rad_hist} it is shown also that a log-normal distribution with a
peak at $\sim$3 pc (dotted line) provides a good fit to data.

The median \reff of the system is $4.2$, with $3.0$ pc standard
deviation. If the three possible sub-populations of SC candidates
(Sect. 3.1) are taken separately, we obtain as median and
standard deviation \reff$=2.9\pm1.5,~3.8\pm1.8,~5.5\pm3.2$ pc for the
blue/intermediate/red peak, respectively (using only the SC candidates
within 1$\sigma$ the colour intervals reported in Fig. \ref{col_site},
the number of objects used are 7, 72, and 18 from blue to red
peak\footnote{We obtain \reff$=3.3\pm1.1,~3.8\pm2.1,~5.3\pm3.5$ pc for
the blue/intermediate/red peak using SC candidates within 2$\sigma$
the colour intervals of Fig. \ref{col_site}.}).  Although the three
median \reff agree within 1$\sigma$, we notice smaller \reff for bluer
SC candidates.  Figure \ref{reffcol} gives more details on this
regard.  In the figure we show the average \reff per colour bin for
both $B-I$ and $B-V$. In spite of the large scatter, it is possible to
observe the presence of a non negligible correlation between the size
and the colour of SC candidates, with red SCs showing on average
larger \reff respect to blue candidates. Such effect has already been
observed in other SC systems \citep[e.g.][]{scheepmaker07}. Again, if
colour differences are associated to age differences, then the
increase of \reff with colour points toward a dynamical evolution of
clusters radii with age, with younger objects being more concentrated
than older ones.

\begin{figure}
\includegraphics[width=\linewidth]{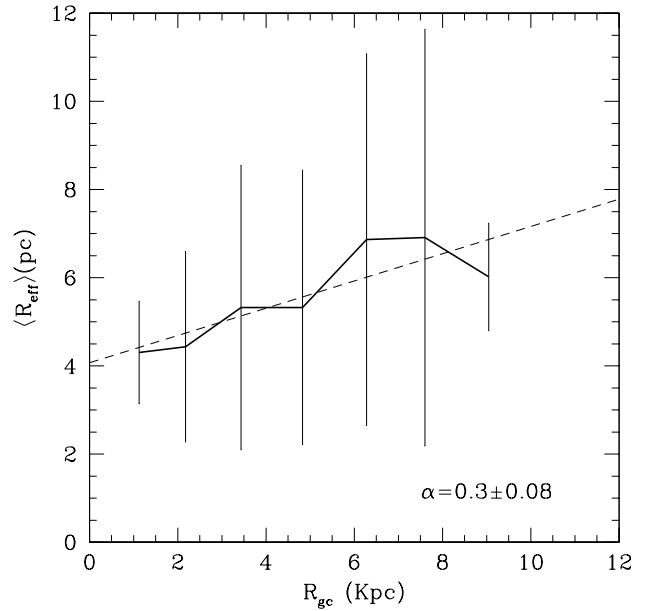}
\caption{Average effective radius versus distance from the galaxy
centre.}
\label{reffrad}
\end{figure}

\begin{figure}
\includegraphics[width=\linewidth]{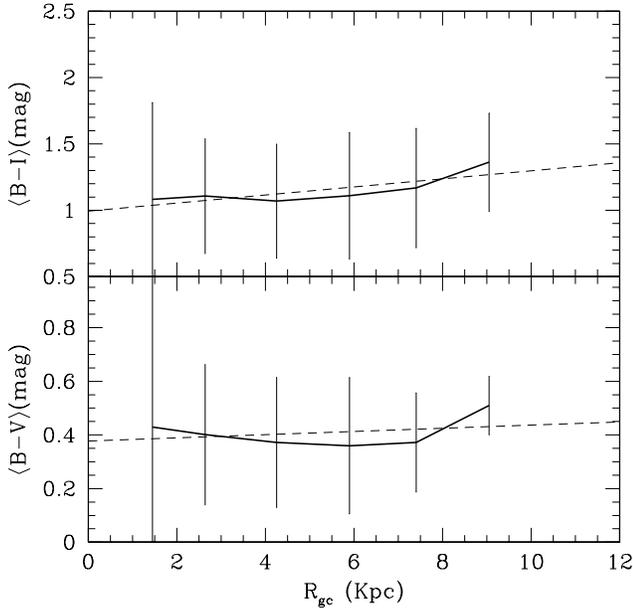}
\caption{Average colour of SC candidates versus galactocentric
distance.}
\label{radcol}
\end{figure}

\begin{figure}
\includegraphics[width=\linewidth]{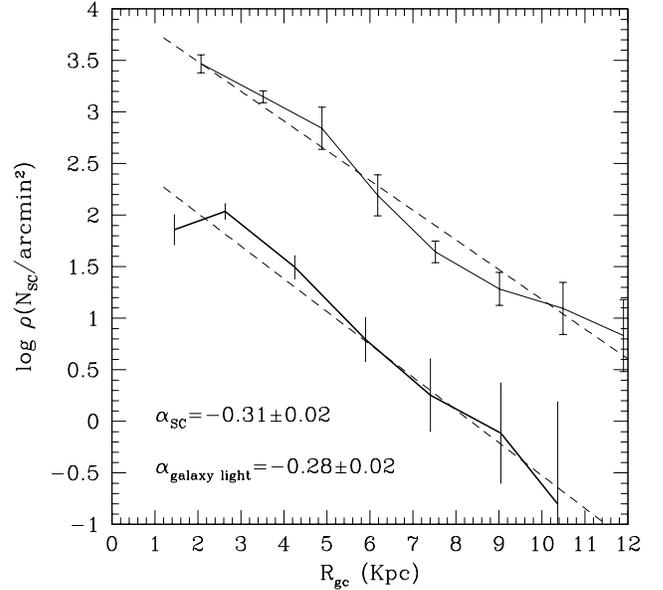}
\caption{Radial density of SC candidates corrected for completeness
(thick solid line). The completeness correction is only significant in
the innermost two bins considered. The upper thin solid line shows the
surface brightness profile of the galaxy shifted vertically by an
arbitrary amount. The slope of a linear fit to both distributions
is also reported in the figure.}
\label{radial_ann}
\end{figure}

The spatial distribution of star clusters provides an additional piece
of information on the properties and origins of the SC system and the
host galaxy.  One possible clue comes from the relation between the
half-light radius and the galactocentric distance $R_{gc}$, shown in
Figure \ref{reffrad}. Also in this case, the spread around the average
\reff for each $R_{gc}$ bin is large, but a correlation over the
entire $R_{gc}$ range seems to be present. A linear fit to the data
provides a slope $\alpha\sim0.30$.  By analysing the SC system of
\object{M 51}, \citet{scheepmaker07} found a shallower slope,
$\alpha=0.12\pm0.02$. However, similarly to \object{M 51}, we find no
correlation, or a very weak one, between the colour of the SC
candidate and $R_{gc}$ (Fig. \ref{radcol}). Finally, the spatial
density of SC candidates versus $R_{gc}$, shown in Figure
\ref{radial_ann}, reveals that the surface brightness profile of the
galaxy (in the figure with an arbitrary shift) agrees with the radial
density profile of SC candidates.  In the figure we omit the
central $\sim$2 kpc as SC candidates in such bright region are
rejected. A linear fit to SC data provides a slope
$\alpha_{SC}=-0.31\pm0.02$, very similar to the
$\alpha_{galaxy~light}=-0.28\pm0.02$ from the surface brightness
profile of the galaxy. Such behaviour, typically observed in late
type galaxies \citep{schweizer96,miller97,larsen99}, might be an
indication that field stars and star clusters have experienced similar
dynamical relaxation processes and, that star clusters at any location
(and age?)  are in dynamical equilibrium with the galaxy potential.

\subsection{Mass function from comparison with models} 

Mass estimates of SCs are sensitive to the age of the cluster,
because the mass luminosity ratio changes with time, due to stellar
evolution, and because of the dynamical evolution of cluster itself
(tidal disruption, evaporation, etc.).  In the approximation of no
dissolution, \citet{gieles07}, using SMC data, find that the minimum
detectable cluster mass scales with age $t$ according to
$M_{min}\propto t^{+0.7}$, or, in other words, the mass detection
limit increases by a factor of $\sim5$ each age dex \citep[see
also][for a similar analysis on clusters in the MW and nearby
galaxies]{boutloukos03}. As discussed in Section 3.1, the present
sample of clusters contains both massive old GCs, which have a peaked
mass function, and young clusters, which form from a power-law mass
function. Therefore, an age dependent mass-completeness affects our
data. To properly constrain the mass of SC candidate, age sensitive
data, like U-band photometry, are essential, and a study of the SC
masses drawn from only BVI data must be considered with extreme
caution.

Keeping in mind these warnings, in this section we tentatively
derive rough mass estimates of SC candidates. To this purpose we use
magnitudes and colours of SC candidates and compare them with stellar
population synthesis models. In particular, we take advantage of the
specific capabilities of the Teramo-SPoT models
\citep{cantiello03,raimondo05}\footnote{Visit the website
www.oa-teramo.inaf.it/spot for details on models, code, and for
downloads.}, to obtain some hints on the mass distribution of the
SCs. Drawing the mass distribution from a comparison of data to SSP
models is a risky exercise, and could be misleading if one does not
consider properly the limitations due to the assumptions done. The
total mass of the SSP strongly depends on the Initial Mass
Function. We assumed a \citet{scalo98} law; a different IMF, like a
\citet{salpeter55} or a \citet{kroupa01} IMF, would affect the total
mass estimates by few tens percent. However, that's rather unimportant
for the rough, and relative analysis that we will carry out here. A
second approximation comes from the fact that the age and mass
associated to each SC candidate depends on the metallicity of the
stellar cluster. For this reason, we have assumed three metallicities,
representative of metal rich, intermediate, and and metal poor stellar
populations, i.e. \feh$\sim$-1.8, -0.7 and 0.0 dex, respectively.
Then, we have analysed the three cases separately.

To derive crude mass estimates of the SC candidates we used two
different approaches, both based on the data to model comparisons:
first we compare SC candidates colour and magnitudes with SSP models
for the three selected \feh separately, then a second estimate of
masses is obtained in the colour-colour plane from median
mass-luminosity ratios.

\begin{figure}
\includegraphics[width=\linewidth]{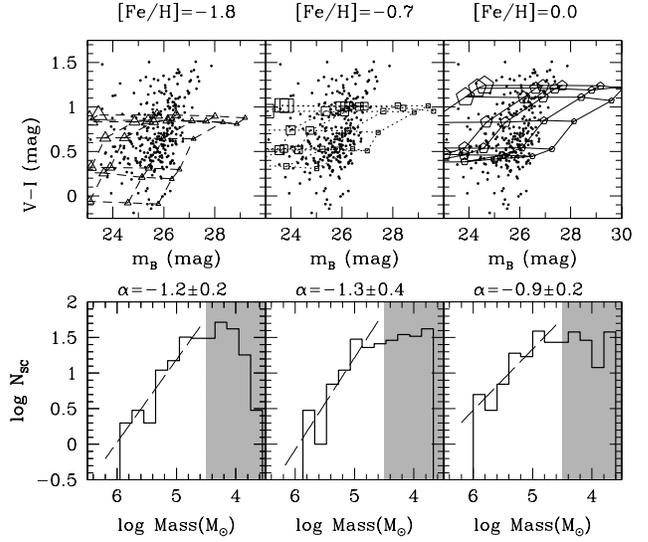}
\caption{Upper panels: magnitude-colour diagrams of SC candidates and
SPoT SSP models for the three reference metallicities ([Fe/H] upper
quotes in each panel). Models with ages 0.03, 0.2, 0.5, 1, 5, 10 and
14 Gyr are connected with a line, older ages have redder
colours. SSPs models with different masses, $M_{tot}\sim 5\cdot
10^3,~1.5\cdot10^4,~6.0\cdot10^4,~1.2\cdot10^5~,~1.0\cdot10^6
M_{\sun}$, are shown with increasing size symbols, depending on the
mass. Models are shifted to the distance modulus of \object{NGC 3370}. Lower
Panels: rough estimates of the Present Mass Functions derived from the
data to model comparison shown in the upper panels. A linear fit to
histograms in the range $4.5 \leq \log (M/M_{\odot}) \leq 6.5$ is
shown with a long-dashed line. The slope of the linear fit is upper
quoted in each panel. The shaded region marks, approximatively, the
mass interval corresponding to a photometric completeness below
$\sim$90\%.}
\label{colors_mods}
\end{figure}

For the specific purpose of this work we have computed SSP models
assuming different total masses, i.e. $5.0 \cdot 10^3,~1.5 \cdot
10^4,~6.0 \cdot 10^3,~1.2 \cdot 10^5,~ 1.0 \cdot 10^6 M_{\odot}$, and
ages 0.03, 0.2, 0.5, 1.0, 5.0, 10.0, 14.0 Gyr, for the three adopted
metal contents. The upper panels in Figure \ref{colors_mods} show a
comparison of data to models for the selected \feh values, all
computed ages and masses are shown.

Assuming a defined \feh, we used the SSP grids shown in Figure
\ref{colors_mods} to constrain the mass of the SC candidate, rejecting
all SC candidates that lie outside all SSP grids by more than $\sim
0.2$ mag.  The lower panels of Figure \ref{colors_mods} show the mass
distributions derived from the data-model comparisons described.

Although we cannot give solid constraints on the present mass function
(PMF) of SCs, it is instructive to study some features of the
magnitude-colour diagrams and PMF derived. First, a few very blue
[$V-I\lsim$ 0.2 mag, $m_{B,0}\gsim25$ mag] SC candidates can only be
associated to a metal poor and young component, since intermediate
\feh and metal rich models do not provide the necessary colour spread
to match these sources. Moreover, the position in the magnitude-colour
diagram of these sources implies low masses, $M\lsim10^4
M_{\odot}$. This is also recognised by the presence of a peak at
$\sim10^4 M_{\odot}$ in the PMF at \feh=-1.8, while the PMF derived
from other \feh are nearly flat below $\sim10^5 M_{\odot}$.  The
expected age-dependent mass completeness \citep{gieles07} is probably
observed in the panels of figure \ref{colors_mods}, too. In fact, only
young SC candidates overlap with SSP models with masses below $\sim1.5
\cdot 10^4 M_{\odot}$, while no SC candidate matches with the grid of
low-mass $t\gsim 1$ Gyr SSP models, no matter what \feh is adopted.
A similar case are the red objects at $V-I\sim1.2$, which can only be
associated to a population of old, metal rich SCs. Although the sharp
drop of the PMFs at masses $\lsim10^4 M_{\odot}$ might be an artifact
of the sample incompleteness, it is worth to note that in all cases a
turn over of the PMF around $\sim10^5 M_{\odot}$ is observed.
Finally, a linear fit to the mass distribution, within the range of
$\sim10^{4.5} M_{\odot}\div\sim10^{6.5} M_{\odot}$, provides a slope
that agrees within 1$\sigma$ uncertainty\footnote{The lower limit to
$\sim10^{4.5} M_{\odot}$ has been chosen to avoid the masses,
i.e. magnitudes, most affected by incompleteness. The B-band
completeness at masses approximatively lower than $\sim10^{4.5}
M_{\odot}$ drops quickly below 90\%.}.

\begin{figure}
\includegraphics[width=\linewidth]{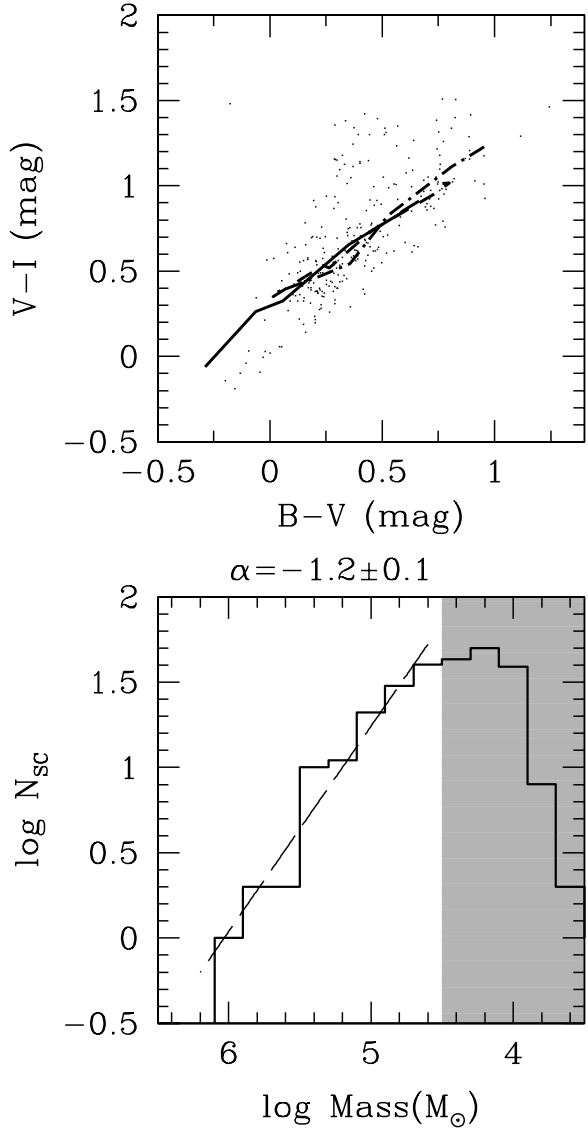}
\caption{Upper panel: SC candidates (dots) compared to SSP colour
predictions from SPoT for \feh=-1.8, -0.7, 0.0 (shown with solid,
dashed and dot-dashed lines, respectively). The models age ranges from
0.03 to 14 Gyr. Lower panel: The PMF derived from the estimate of
median $M/L_V$ ratios of SC candidates.}
\label{ml_blake}
\end{figure}

As an additional test we have obtained masses by using the average
mass-luminosity ($M/L_V$) ratios derived from the $B-V$ versus $V-I$
colour-colour diagram, then converting V-band magnitudes to masses.
More in details, we have constrained the mean $M/L_V$ of the SC
candidates by the comparison of SC colours with SSP models in a broad
range of ages ($0.03 \leq t(Gyr) \leq 14.0$) and [Fe/H] from -2.3 to
+0.3 dex. We set a default 0.25 ``colour distance'' to the SSP model
for accepting its $M/L_V$ ratio\footnote{We made several tests using
different ranges (from 0.1 to 0.5 mag, also releasing the matching
between colours), without any substantial change in the PMF
slope.}. The upper panel of Figure \ref{ml_blake} shows the
colour-colour diagram of SC candidates, and the three reference SSP
models from SPoT. The median $M/L_V$ obtained from the data-to-models
comparison was then used, together with visual magnitude of the SC
candidate, to estimate the mass. The main difference with this new
approach is that no a priori assumption is done on the \feh
content. Moreover, in the previous approach, at fixed \feh the mass
and age of the SC were virtually present in the comparisons shown in
Figure \ref{colors_mods} (upper panels), while in this case mass comes
from the statistical analysis of the $M/L_V$ of SSP models within
large age and \feh ranges. The PMF derived using $M/L_V$ ratios is
shown in Figure \ref{ml_blake}. Our main interest, here, is to
emphasise that the two approaches used to derive the PMF provide
similar results concerning the power law {\it profile}.

The PMF in spiral and irregular galaxies is generally well fit by a power-law,
although the fit is limited to the young and most massive component of
clusters. The typical value for power law exponent is $\alpha\sim-2.0$
\citep[e.g.][]{zhang99,bik03}. Similarly, the high mass tail of
Galactic OCs has a slope $\alpha=-2.04 \pm 0.11$, with no selection on
clusters age, thus including objects from few Myr up to several Gyr
\citep{battinelli94}. 

Such numbers agree with the slopes obtained from the power-law fit to
data (shown in the Figures \ref{colors_mods}-\ref{ml_blake}), ranging
between $-1.3\pm0.4$ and $-0.9\pm 0.2$, and corresponding to a power
law index $\alpha = -2.3\div -1.9$ in the form $dN_{SC}/dM\propto
M^{\alpha}$.

Recently, \citet{larsen09} proposed that the power-law fit to the MF
of SCs might be due to the limited mass range of observational data,
and suggested that a functional form of the type of a Schechter law -
characterised by a power law and an exponential decline after some
characteristic mass $M_c\sim 2.1 \times 10^5 M_{\sun}$ - approximates
quite well some of the best studied SC systems, like in \object{M 51} and
\object{LMC}. However, \citeauthor{larsen09} also warns that it is unlikely
that such functional form is universal, quoting the cases of some
merger or starburst galaxies as examples where significantly higher
$M_c$ values are required. As a non-interacting spiral, \object{NGC 3370}
should fall in the sample of galaxies with a PMF well described by a
Schechter form. Assuming a Schecter function as in \citet{larsen09}, a
cut-off mass of the order of $M_c \sim 10^6 M_{\sun}$ would be
required for the approximate PMF derived here. Such high $M_c$ value
has been fit to the young SCs in the Antennae \citep{jordan07}, but it
is highly unlikely in spirals, being probably a result of the sample
incompleteness at low masses, and the rough mass estimates used here.

\subsection{Globular Cluster candidates}
\label{sect:gc}
\begin{figure}
\includegraphics[width=\linewidth]{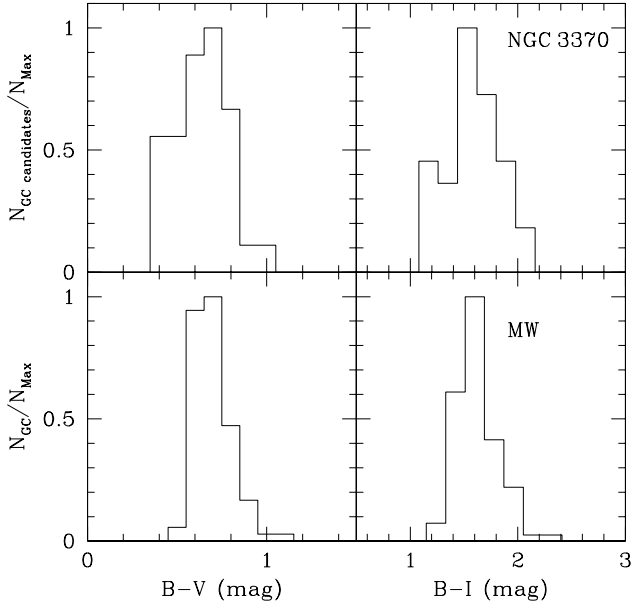}
\caption{Colour histograms of GC candidates in \object{NGC 3370} (upper
panels) and of bright GGCs (lower panels). The histograms are
normalised to the bin with maximum amplitude.}
\label{gccol}
\end{figure}

The colour distributions of \object{NGC 3370} clusters, and the comparisons
with GGCs, as well as with the cluster system in other galaxies, shown
in previous sections, has evidenced the presence of a red system of
objects with properties remarkably similar to old globular
clusters. In this section we provide a more detailed study on such GC
candidates. To select a sample of candidate GCs we consider only
sources with $V-I$ matching with the GC system in the Galaxy, in
addition sources in the brightest/bluest part of the disk are rejected
to avoid contamination ($R_{gc~min} \geq 50\arcsec$, $\mu_{V}\geq
24.0~mag/arcsec^2$). With such criteria we end up with a list of 35 GC
candidates. Figure \ref{gccol} shows the colour histograms (normalised
to the maximum bin amplitude) for GCs in \object{NGC 3370} and, for
comparison, for GGCs brighter than $M_V \gsim -5$ \citep[GGC data
from][]{harris96}.  The median colours are $B-V=0.62\pm0.15$ mag
($0.68\pm 0.11$ mag), and $B-I=1.60\pm0.27$ mag ($1.58\pm 0.19$ mag)
for \object{NGC 3370} (MW) star clusters. The median \reff is $\sim 5.5$ with
3.9 pc standard deviation.  Having in mind the warning of the small
sample of GC selected, it is possible to associate the similarity
between the $B-V$ and $B-I$ colour distributions in both galaxies to
similar age/metallicity properties of the two GC systems. However,
some differences seems to be recognised between the two
galaxies. First, the colour distributions in \object{NGC 3370} do not appear
so sharp as in the case of GGC, though this might be a consequence of
larger photometric uncertainties. Second, both \object{NGC 3370} distributions
show a tail of GC at bluer colours with respect to the MW - more
evident in $B-I$. If the apparent secondary peak is interpreted as a
metallicity difference then, using the $B-I$ colour-metallicity
relation by \citet{harris06} derived from GGCs, we find that the main
peak corresponds to a metallicity [Fe/H]$\sim-1.5$ dex (as for the
Galaxy), while the \feh for objects in the blue tail is $\sim-2.3$
dex. However, we must emphasise that several doubts have been raised
in the last few years against the simple interpretation of colour
bimodalities in GC systems as metallicity bimodalities.  Non-linear
metallicity-colour transformations, in fact, have been proved to be
effective in generating multimodal colour distributions from unimodal
\feh distributions \citep[see][and references therein]{cantiello07d}.

One other useful quantity related to the GC system is the specific
frequency, $S_N$, i.e. the number of GC normalised to the host galaxy
luminosity. This number provides a useful quantity related to the
formation mechanisms of the galaxy \citep[see][and references
therein]{brodie06}. To properly estimate $S_N$, however, the total
number of GC needs to be used. As a consequence of the sample
incompleteness we cannot reliably constrain the specific frequency,
but only provide an estimate to the lower limit of $S_N$. In
particular, we make the following assumptions: $i)$ we use the results
by, e.g., \citet{rhode03} showing that the surface density of GC in
spirals follows a power law profile; $ii)$ a total apparent corrected
magnitude $m_{V,0}\sim 11.7$ mag for \object{NGC 3370} from
Hyperleda\footnote{http://leda.univ-lyon1.fr} is adopted, and $iii)$ a
the slope for the density profile of $\log \rho_{GC} \sim -1.5 \log
R_{gc}$ is fixed from the compilation of \citet{kp97}, for a total
galaxy magnitude $M_V\sim21.5$ mag. By constraining the intercept of
the density profile from the number of GCs within
$50\arcsec\div80\arcsec$ - corrected for photometric completeness - we
find a lower limit of $S_N\sim 0.8$, to be compared with $S_N\geq 0.5$
for the GC systems on other spirals \citep{kp97,rhode03,chandar04}.

\section{Summary}

Deep optical VI and shallower B band observations of the face on
spiral galaxy \object{NGC 3370}, drawn from the ACS/HST archive, have been
used to study the SC system in the galaxy. The characteristics of the
imaging data available allowed to analyse the photometry of SC
candidates, and their spatial properties. The final list of SC
candidates consists of 277 objects, selected according to maximum
colour, magnitude, size and position within the galaxy.

The analysis to the sample of SC candidates leads to the following
conclusions:

(1) the colour distributions show that the SC population is composed
    by different subpopulations. In particular, the differences are
    related to ages, with the main sub-populations represented by $i$)
    a red system of likely GC candidates, $ii$) a dominant peak of
    sources with properties similar to intermediate age clusters
    (similar to OCs with ages from a few 100 Myrs, to few Gyrs), and
    $iii$) a blue tail of SCs, possibly with ages below $\sim$100
    Myr. Such complex age-structure of the SC system is somehow
    expected in a Sc-type spiral, actively star-forming (as also
    testifies the amount of Cepheids observed in this galaxy);
    
(2) the comparison of SC candidates colours - selected with no a
    priori assumption on the range of colours allowed - with SSP
    models predictions within a large range of ages (30 Myr to 14 Gyr)
    and \feh (-2.3 to +0.3 dex) shows a good matching between models
    and data. A significant exception to this is represented by a
    sample of $\sim$ 20 SC candidates, which systematically lie
    outside the region of SSP models, showing an excess of flux at
    longer wavelength (I-band magnitude).  Such SC candidates do not
    show remarkable peculiarities in terms of spatial distributions or
    morphology. One possible origin of their behaviour is that they
    are typical intermediate age SCs (300 Myr to 2 Gyr).  At these
    ages, the stochastic presence of TP-AGB stars has a dramatic
    impact on the integrated light. In particular, TP-AGBs dominate
    the long-wavelength emission of the stellar system \citep[see,
    e.g.,][and references therein]{maraston05,marigo08}, thus the
    stochastic presence of TP-AGBs might cause the observed position
    of these SCs in the colour-colour diagram \citep{raimondo09};

(3) the median \reff of the system is $4.2$ with 3.0 pc standard
    deviation. The \reff distribution shows a power law drop above
    \reff$\sim 3.0$ pc. The exponent of the power law,
    $\eta=2.35\pm0.22$, appears normal with respect to other similar
    galaxies. Moreover, a log-normal fit to the distribution of \reff
    seems to show a turn-over around $\sim$3 pc, although the
    completeness of the data does not allow to give more robust
    conclusions;

(4) the \reff shows non-negligible correlations both with the colour
    of the SC candidate, and with the galactocentric distance
    $R_{gc}$. If colours differences are associated to age
    differences, as suggested in (1), these results point toward an
    age evolution of \reff, with older SCs having larger
    \reff. Furthermore, the spatial density of clusters, which appears
    similar to the surface brightness profile of the galaxy, might
    well be a consequence of a common dynamical evolution of stars and
    star clusters in the gravitational potential of the galaxy;

(5) the LF of SC candidates shows a continuous increase down to
    magnitudes where the SC candidate sample is fairly unaffected by
    incompleteness. Such LF is well described by a power law with
    exponent slightly below $\sim-2.0$ in all bands. This
    functional form agrees to that of other SC systems in spiral or
    irregular galaxies. A comparison of the maximum SC magnitude
    with the SFR per unit area shows that the correlation between
    these quantities is like in other spirals. Moreover, some
    properties of the LFs presented here (steepening of the LF towards
    brighter magnitudes, and steeper slopes at longer wavelengths)
    seem to support the idea that the SCLF is originated by a
    truncated mass function, though sample incompleteness hampers more
    detailed conclusions;

(6) by comparing data to SSP models, we tentatively derived a crude
    estimate of the Present Mass Function of the SC system. Since our
    data suffer for the lack of an age-sensitive indicator, the PMF
    derived must be taken carefully as a best guess with available
    data. With this warning in mind, we find that a power law fit to
    the interval of more massive SCs (i.e. objects likely less
    affected by the incompleteness) provides an
    $\alpha\sim-2.3\div-1.9$, in agreement to what can be found in
    literature for similar galaxies. Moreover, the PMF seems to show a
    turn-over around masses of the order of $\sim10^5M_{\odot}$,
    though this is possibly due to the sample incompleteness at
    lower masses;

(7) we analysed the properties of a sample of GC candidates, selected
    on the base of their $V-I$ colour, and their positions with
    respect to the centre of the galaxy. The number of GC candidates
    selected is 35, these show colour distributions similar to GGCs,
    though less peaked around a median value (probably due to the
    larger photometric uncertainties), and with a relatively more
    populated tail towards bluer colours. The peak of the colour
    distribution, converted to metallicity using GGCs data, gives
    \feh$\sim-1.5$, similar to GGCs;

(8) due to the complex issue of completeness in these data, we cannot
    provide a reliable estimate to the specific frequency of
    GCs. However, taking advantage of the known approximative
    behaviour of the surface density profile of GC in spirals, we
    estimate a lower limit $S_N>0.8$.

The high resolution of ACS/WFC allowed us to carry on a refined study
of the structural and photometric properties of the SC system in
\object{NGC 3370}. These data are essential to provide further constraints to
the evolution of stars and stellar systems in this galaxy. On the
other hand, the lack of age-sensitive indicators, like
shorter-wavelength ($\sim$ U-band) photometry, is a serious drawback
for a detailed study of ages and masses of the SC system in this
galaxy. The restoration of ACS and the installation of WFC3 on the
HST, after SM4, will certainly help to fill the present lack of data.

\begin{acknowledgements}
It is a pleasure to thank Gabriella Raimondo for several scientific
discussions and for her irreplaceable support in computing theoretical
models. We are also grateful to S\"oren Larsen for his fruitful
comments and suggestions on data analysis. Part of this work was
supported by PRIN-INAF 2006 (PI G. Clementini). We would thank
the referee, Dr. Mark Gieles, for valuable remarks and suggestions.
\end{acknowledgements}


\bibliographystyle{aa}
\bibliography{cantiello_apr09}

\end{document}